\documentclass[12pt,draftclsnofoot,onecolumn]{IEEEtran}
\usepackage{graphicx} 
\usepackage{subfigure}
\usepackage{color}
\usepackage{algorithm}
\usepackage{amsmath}
\usepackage{amsfonts}
\usepackage{tabularx}
\usepackage{multirow}
\usepackage{caption}
\title{\LARGE A Comprehensive Comparison between Terahertz and Optical Wireless Communications}
\author{\large \IEEEauthorblockN{Mingqing Liu,~Hossein Kazemi,  \emph{Member, IEEE,} Majid Safari, \emph{Senior Member, IEEE,}~Iman Tavakkolnia, \emph{Member, IEEE,}~and~Harald Haas,  \emph{Fellow, IEEE}}
\thanks{This paper was presented in part at 2024 IEEE Global Communications Conference~\cite{liu2024energy}.}
\thanks{This work is financially supported by the UK Government funded project (REASON) under the Future Open Networks Research Challenge (FONRC) sponsored by the Department of Science Innovation and Technology (DSIT) and in part by the Engineering and Physical Sciences Research Council (EPSRC) under grant EP/X040569/1 `Future Communications Hub in All-Spectrum Connectivity'.
Mingqing Liu,~Hossein Kazemi,~Iman Tavakkolnia,~and~Harald Haas are working with LiFi Research and Development Centre, Electrical Engineering Division, University of Cambridge, Cambridge, UK (email:\{ml2176, hk572, it360, huh21\}@cam.ac.uk). Majid Safari is working with Institute for Imaging, Data, and Communications, School of Engineering, University of Edinburgh, Edinburgh, UK (email: majid.safari@ed.ac.uk).}

}

\begin{document}

\maketitle
\begin{abstract}
    This paper presents a comprehensive quantitative comparison between Terahertz (THz) communication (TeraCom) and optical wireless communication (OWC) technologies, focusing on both indoor and outdoor environments. We propose a comparison method for TeraCom and vertical-cavity surface-emitting laser (VCSEL)-based OWC in indoor scenarios, incorporating misalignment effects by modeling the THz antenna radiation pattern within a multi-ray THz channel model and using a Gaussian beam model for VCSEL-based OWC. Unified beamwidth parameters allow for a detailed analysis of misalignment impact on both systems. Furthermore, we develop power consumption models for each technology, integrating key parameters such as THz phase noise, VCSEL non-linearities, and photodetector bandwidth-area tradeoffs. These models enable an in-depth analysis of energy efficiency in indoor environments, including multi-transmitter coverage scenarios. For outdoor scenarios, we summarize existing stochastic channel models addressing path loss, pointing errors, and small-scale fading for free space optics (FSO) and THz links. We then apply these models to unmanned aerial vehicle (UAV) applications to assess performance in dynamic conditions. Our results provide critical insights into the suitability of each technology for various deployment scenarios.
\end{abstract}

\section{Introduction}
The rapid growth in demand for high-speed, high-capacity wireless communication has driven interest in Terahertz (THz) and optical wireless communication (OWC) technologies, each offering unique strengths for short- and long-distance links~\cite{THzfirst,FSOfirst}. THz links, operating in the high-frequency spectrum from 0.1 THz to 1 THz, and OWC, leveraging visible and infrared light, offer promising alternatives to traditional RF technologies, especially in scenarios where bandwidth or interference constraints are critical~\cite{THzoverviewRasp}. From existing experimental demonstrations, both technologies are capable of supporting multi-kilometer Tbit/s wireless data transmission~\cite{THztestbed,FSOtestbed}. Still, comparing these technologies is essential to identify their respective strengths and limitations in addressing diverse communication needs. On the one hand, the comparison of transmission characteristics of THz and optical wireless technologies enables us to find appropriate application scenarios for each technology. On the other hand, the performance gap and corresponding impact factors of THz and optical wireless in a common application scenario still need demonstration to date. Thus, this manuscript aims to provide a comprehensive comparison of these two technologies.

Most existing comparative studies between THz and OWC technologies are primarily qualitative, providing general insights into their characteristics but lacking quantitative performance assessments~\cite{compare,comparethree,comparetwo}. However, quantitative analysis is crucial for understanding the practical viability of these technologies under specific conditions, such as power efficiency, alignment precision, and environmental impacts. Quantitative comparisons can provide more concrete guidance for technology selection and optimization in real-world deployments. OWC can be categorized into short-range indoor OWC, often referred to as LiFi, and long-distance free space optics (FSO)~\cite{OWCoverview}. Similarly, THz links are divided into indoor and outdoor applications to address distinct propagation characteristics in each scenario~\cite{THzoverviewRasp,THzoverview2}. Thus, this paper conducts quantitative analyses for both indoor and outdoor environments, enabling a more comprehensive comparison of Terahertz communication (TeraCom) and OWC performance under varying conditions.

In indoor settings, vertical-cavity surface-emitting laser (VCSEL)-based OWC is regarded as a potential “LiFi 2.0” solution~\cite{LIFI2}, while electronics-based THz technology offers a cost-effective and mature alternative compared to photonic- and plasmonic-based THz technologies~\cite{electronicsBetter}. Thus, we focus our indoor comparison on these two systems. To mitigate the significant path loss at high frequencies, both THz and OWC systems require directional transmission with energy-concentrating beams, making misalignment analysis essential. Additionally, energy efficiency is a critical factor in indoor scenarios, leading us to establish channel models that account for misalignment effects and develop energy consumption models for a detailed numerical analysis~\cite{VCSELEE,THzEE1,THzEE2}. Through quantitative analysis, we demonstrate that VCSEL-based OWC outperforms TeraCom in energy efficiency but is more sensitive to transmission distance and bandwidth limitations. Consequently, VCSEL-based OWC requires a more focused beam to mitigate these challenges effectively.

For outdoor environments, extensive research has been conducted on stochastic channel models for FSO~\cite{FSOstochannel1,FSOstochannel2} and THz links~\cite{THzchannelFTR,generalTHzchannel}. These models describe essential characteristics, including THz absorption properties and the sensitivity of FSO to turbulence and weather conditions, as well as the pointing error characteristics of both channels. Summarizing and analyzing these models through numerical comparisons can reveal key similarities and differences between FSO and THz links. Furthermore, applying these models to unmanned aerial vehicle (UAV) scenarios, which are particularly sensitive to pointing errors, can provide valuable insights into the suitability of each technology for mobile outdoor applications, helping identify the optimal solution for dynamic and challenging environments~\cite{THzUAV,FSOUAV,FSOUAV2}. Numerical results by the comparative analysis reveal the impact of weather conditions, absorption, and pointing errors on both links. Besides, FSO is more sensitive to transmission distance and misalignment, making it less suitable for UAV-to-UAV applications compared to THz links.

The contribution of this manuscript is summarized as
\begin{itemize}
\item[1)] We propose a method to compare TeraCom and VCSEL-based OWC in an indoor environment with the misalignment effect. For TeraCom, we incorporate the antenna radiation pattern with pointing directions into a multi-ray THz channel model, capturing the effects of directional beam propagation. For VCSEL-based OWC, we employ a Gaussian beam model that accounts for geometric misalignment. By unifying the beamwidth parameters of the THz and optical antennas, this approach enables a quantitative analysis of each system's performance and the impact of misalignment on signal quality.
\item[2)] We develop power consumption models for TeraCom and VCSEL-based OWC, incorporating key factors such as cascaded component power efficiency and phase noise for THz systems, and non-linear conversion effects of VCSELs along with the bandwidth-area tradeoff of photodetectors. Using these analytical models alongside channel models, we perform a numerical analysis to compare the energy efficiency of both systems. Additionally, we extend this analysis to an indoor environment with multiple transmitters to evaluate coverage and service quality.
\item[3)] We summarize existing stochastic channel models for FSO and THz links, covering path loss, pointing error, and small-scale fading, to assess link performance in outdoor environments with various environmental conditions. In particular, we compare two different pointing error models for THz channels. Finally, we apply these channel models to UAV applications to evaluate link performance in dynamic scenarios.
\end{itemize}

The remainder of this manuscript is organized as follows. Section II provides a comparative overview of THz and OWC technologies, detailing their respective key elements. In Section III, we present analytical models for comparing TeraCom and VCSEL-based OWC in indoor environments, addressing the effects of misalignment, energy efficiency, and multi-transmitter coverage. Numerical analysis is then conducted to show these comparisons. Section IV explores FSO and THz link performance using various stochastic channel models, with an application to UAV scenarios for performance evaluation. Conclusions are drawn in Section VI. 

\section{Comparative overview of TeraCom and OWC}

This section provides a comparative overview of key elements, i.e., front-end devices, antenna and detection technologies, propagation characteristics, mobility support, and digital backend, in THz and OWC technologies, highlighting their respective advantages and challenges.

TeraCom technologies utilize various front-end devices. CMOS-based electronic THz devices balance transmitted power and energy efficiency~\cite{elecTHz}, while semiconductor-based THz sources offer higher power and efficiency but face higher costs and integration challenges~\cite{semiconductorTHz}. Photonics-based THz devices provide low phase noise and high speed but suffer from low power efficiency and implementation difficulties~\cite{photoTHz}. Additionally, plasmonic devices show potential for high power and bandwidth but are hindered by high cost and technical immaturity~\cite{plasmTHz}. TeraCom primarily employs horn antennas with gains up to 55 dBi, but their bulkiness limits mobile applications. THz antenna arrays are still under development, particularly at higher frequencies, where precise beamforming remains challenging~\cite{THzarray}.

While the hardware for TeraCom systems still requires further effort, OWC technologies primarily employ publicly accessible light sources, such as LEDs and laser diodes (LDs) in the transmitter, and photodetectors (PDs) and photovoltaic (PV) cells in the receiver~\cite{FSOfirst}. LEDs are widely available and low-cost but are limited in bandwidth and output light intensity, which can restrict data rates and coverage. LDs, especially vertical-cavity surface-emitting lasers (VCSELs), provide higher efficiency and bandwidth, though they require strict eye-safety regulations~\cite{VCSELmodel2,VCSELtrans}. Advances in micro-LED and micro-VCSEL arrays further enhance bandwidth capacity, suggesting their potential for high-speed OWC applications. PDs and PV cells offer high response speeds and can be adapted for simultaneous lightwave information and power transfer (SLIPT)~\cite{OWCoverview}. However, there is a tradeoff between bandwidth and detection area, affecting the field of view (FOV) and signal sensitivity. Camera-based receivers, while offering larger FOVs, tend to operate at lower speeds and thus are limited to lower-data-rate applications.

Propagation characteristics play a crucial role in system performance. Indoor TeraCom propagation features low absorption but suffers from medium scattering and high blockage sensitivity, with ray-tracing models commonly used to analyze multipath effects~\cite{multiRay}. Outdoors, THz signals experience high molecular absorption and low scattering, affected by air composition and humidity~\cite{THzchannelFTR,generalTHzchannel}. Stochastic models are applied to address pointing errors and small-scale fading. Conversely, indoor OWC propagation varies by source type: LED-based systems exhibit larger divergence and non-line-of-sight (NLOS) reflections~\cite{LEDNLOS}, whereas LD-based systems provide more concentrated beams with negligible multipath reflections~\cite{VCSELtrans}. Deterministic analytical models are used for indoor OWC propagation. Outdoor OWC experiences higher scattering and absorption, further affected by weather conditions and turbulence. Like THz, stochastic models are employed to address pointing errors and fading effects~\cite{FSOstochannel1,FSOstochannel2}.

Mobility solutions in THz and OWC are at different maturity levels. THz systems are exploring beamforming and multiple-input-multiple-output (MIMO) techniques, but practical deployment remains challenging due to hardware and alignment complexities~\cite{monnai2023terahertz}. In OWC, optical beam steering and MIMO techniques enable mobility~\cite{VCSELtrans}, but beam steering is costly, and expanding receiver FOV requires further development~\cite{PDtradeoff}. Besides, both systems share challenges in digital backend processing~\cite{THzoverviewRasp}. As required data rates increase, the size, cost, and thermal requirements of digital-to-analog converters (DACs) and analog-to-digital converters (ADCs) become bottlenecks, necessitating efficient, low-power signal processing.

\section{Indoor applications: TeraCom vs. VCSEL-based OWC}
This section introduces a typical VCSEL-based OWC system for indoor scenarios to compare with the TeraCom system. The channel gain models of the two systems for single-input-single output (SISO) links are presented first. After specifying the noises considered in the systems, e.g., phase noises in TeraCom, the signal to noise ratio (SNR) of the two systems is formulated. Besides, with typical system designs, the consumption factor is adopted to indicate the systems' energy efficiency. Based on models built for SISO links, distributed AP based-networks are considered to analyze the system performance on data rate as well as energy efficiency. At last, we conduct the numerical analysis to compare the two systems quantitively.
\subsection{SISO link: VCSEL-based OWC}
\subsubsection{Channel model with misalignment} For laser systems, the Gaussian beam propagation principle is widely adopted to express diffraction loss while the laser propagates in the free space. In~\cite{VCSELtrans}, a Gaussian beam propagation-based unified misalignment channel model is built for VCSEL-based OWC, where the displacement error along $x,y-$axis denoted by $x_{\rm dis}$ and $y_{\rm dis}$, azimuth and elevation angles of orientation error at Tx and Rx side denoted by $\phi_{\rm a}^{\rm t}$, $\phi_{\rm e}^{\rm t}$, $\phi_{\rm a}^{\rm r}$, and $\phi_{\rm e}^{\rm r}$ are included. The channel gain is expressed as follows~\cite{VCSELtrans}
\begin{equation}
    H_{\rm owc}=\iint_{(x, y) \in \mathcal{A}} \frac{2}{\pi w^2(L)} \exp \left(-\frac{2 \rho^2(x, y)}{w^2(L)}\right)  {\rm d} x {\rm d} y,
\end{equation}
where $\mathcal{A}$ is the receiving PD surface. $w(L)$ is the beam waist at $z=L$ and $w^2(L)$ is defined by
\begin{equation}\left\{
   \begin{aligned}
       &w^2(L) = w_0^2\left[1+z_{\rm R}^{-2} (c_1+xc_2+yc_3-c_4)^2\right]\\
       &c_1:=L\cos\phi_{\rm e}^{\rm t}\cos\phi_{\rm a}^{\rm t},\quad c_2:=\cos\phi_{\rm e}^{\rm t}\sin(\phi_{\rm a}^{\rm t}-\phi_{\rm a}^{\rm r})\\
       &c_3:=\sin\phi_{\rm e}^{\rm r}\cos\phi_{\rm e}^{\rm t}\cos(\phi_{\rm a}^{\rm t}-\phi_{\rm a}^{\rm r})+\cos\phi_{\rm e}^{\rm r}\sin\phi_{\rm e}^{\rm t}\\
       &c_4:=x_{\rm dis}\cos\phi_{\rm e}^{\rm t}\sin\phi_{\rm a}^{\rm t}-y_{\rm dis}\sin\phi_{\rm e}^{\rm t}
   \end{aligned}\right.,
\end{equation}
where $w_0$ is the primary beam waist radius and $z_{\rm R} = {\pi w_0^2}/{\lambda}$ with $\lambda$ indicating the beam wavelength. Besides, $\rho(x, y)$ is the
Euclidean distance of the point $(x,y)$ from the center of the beam spot, and $\rho^2(x, y)$ is expressed as
\begin{equation}
\begin{aligned}
\rho^2(x, y) = &\left[ L-x\sin\phi_{\rm a}^{\rm r}+y\cos\phi_{\rm a}^{\rm r}\sin\phi_{\rm e}^{\rm r}\right]^2+\left[ x\cos\phi_{\rm a}^{\rm r}+y\sin\phi_{\rm a}^{\rm r}\sin\phi_{\rm e}^{\rm r}-x_{\rm dis}\right]^2+\\
& (y\cos\phi_{\rm e}^{\rm r}-y_{\rm dis})^2+(c_1+xc_2+yc_3-c_4)^2.
\end{aligned}
\end{equation}

\subsubsection{Channel gain enhancement}The channel gain is closely related to the primary beam waist radius $w_0$. That is smaller $w_0$ leads to large beam divergence, and thus larger transmission loss. Generally, an optical element such as a lens, is embedded in front of the VCSEL transmitter to enlarge the beam waist, achieving beam focusing to enhance the received signal power. Denoting $G_{\rm lens}$ as the magnification factor of the lens, which is defined as the ratio of the transformed beam waist to the original beam waist, the transformed primary beam waist $w_0'=G_{\rm lens} w_0$.
There is also a tradeoff between the magnification factor and beam divergence. Adopting the half-power beam divergence (HPBD) full angle as a metric, this tradeoff is formulated as~\cite{divergence}
\begin{equation}
    \theta^{\rm Mag}=\sqrt{2\ln 2}\frac{\lambda}{\pi G_{\rm lens} w_0}.
\end{equation}
$\theta^{\rm Mag}$ is the HPBD full angle after lens from the VCSEL-based system.

Besides, the PD area impacts the channel gain. However, there is a tradeoff between the bandwidth and area of a PD, which is expressed as
\begin{equation}
    B = \left(\sqrt{\frac{4\pi \varepsilon_0\varepsilon_{\rm r}R_{\rm L}}{0.44 v_{\rm s}}A_{\rm PD}}\right)^{-1},
\end{equation}
where $A_{\rm PD}$ indicates the PD area and $R_{\rm L}$ is the load resistance of the detector. Permittivity in vacuum $\varepsilon_0=8.854\times 10^{12}$F/m, relative
permittivity of the semiconductor $\varepsilon_{\rm r}=12$, carrier saturation velocity $v_{\rm s}=1\times10^5$m/s are taken from \cite{PDdesign}. One of the existing approaches utilizes compound parabolic concentrators (CPCs) to increase the effective area of a PD as~\cite{PDtradeoff}
\begin{equation}
    \hat{A}_{\rm PD} = \frac{\pi }{4}G_{\rm CPC} A_{\rm PD},
\end{equation}
where $G_{\rm CPC}$ is the CPC gain limited by the acceptance angle $\theta_{\rm CPC}$ for incident beam as
\begin{equation}
    G_{\rm CPC} = \frac{1}{\sin^2{\theta}_{\rm CPC}}.
\end{equation}

\subsubsection{Signal-to-noise ratio} A typical VCSEL-based OWC system architecture contains a transmitter and a receiver. At the transmitter, a bias-Tee supplying power for VCSEL, a VCSEL, and optical elements to focus the beam are included; while at the receiver, the detector, i.e., PD, and trans-impedance amplifier (TIA) after it are contained. 
To formulate the SNR of the VCSEL-based OWC systems, shot noise, thermal noise, and laser relative intensity noise are considered. The noise variance of VCSEL-based OWC is given by
\begin{equation}
    \sigma^2 = \frac{4kTBF_{\rm n}}{R_{\rm L}}+2qBR_{\rm PD}H_{\rm owc}P_{\rm t}+BN_{\rm RI}(R_{\rm PD}H_{\rm owc}P_{\rm t})^2,
\end{equation}
where $q$ is the elementary charge, $B$ is the system bandwidth, $R_{\rm PD}$ is the responsivity of PD, $P_{\rm t}$ is the transmitted laser beam power from VCSEL, $k$ is Boltzmann constant, $T$ is the temperature in Kelvin, $F_{\rm n}$ is the noise figure of the trans-impedance amplifier (TIA), $N_{\rm RI}$ is the power spectral density of the VCSEL's relative intensity noise defined
as the mean square of instantaneous power fluctuations divided
by the squared average power of the laser source.

Meanwhile, the optical direct-current (DC) bias power is required to comply with the non-negativity constraint and guarantee better communication performance. For direct current optical orthogonal frequency division multiplexing (DCO-OFDM) scheme, average signal power $P_{\rm sig} = \frac{1}{9}P_{\rm t}^2$ is chosen to discard the clipping noise effectively. Then, the SNR of the VCSEL-based OWC system is derived as
\begin{equation}
    \gamma_{\rm owc} = \frac{R_{\rm PD}^2H_{\rm owc}^2P_{\rm sig}}{\sigma^2}.
    \label{eq:owcsnr}
\end{equation}

\subsubsection{Consumption factor}
At the VCSEL-based OWC transmitter, DC power and signal are input into bias-Tee and then injected into VCSEL to be coveted to beam power. The photoelectric conversion is non-linear. Light intensity-current-voltage (LIV) curves are used to represent the input-output characteristics of VCSELS, which can be obtained through measurements. In~\cite{VCSELmodel2}, the relationship between the output optical power $P_{\rm out}$ and input drive current $I_{\rm in}$ of VCSEL is 
\begin{equation}
    P_{\rm out} = \left\{\begin{aligned}
        \frac{\eta (I_{\rm in}-I_{\rm th})}{\left[1+\left(\frac{\eta (I_{\rm in}-I_{\rm th})}{P_{\rm sat}}\right)^{2N_{\gamma}}\right]^{\frac{1}{2N_{\gamma}}}}, &\quad I_{\rm in}\ge I_{\rm th}\\
        0, &\quad I_{\rm in}<I_{\rm th}
    \end{aligned},\right.
    \label{eq:VCSELmodel2}
\end{equation}
where $P_{\rm sat}$ is the saturation power, $I_{\rm th}$ is the threshold current, $\eta$ is the conversion efficiency in W/A, and $N_{\gamma}$ is a parameter describing the curve fineness~\cite{VCSELmodel2}. These parameters are obtained through measurement and data fitting. We denote $\eta_2(\cdot)$ as an operator to indicate the non-linearity function in Eq.~\ref{eq:VCSELmodel2}. Given input bias voltage $V_{\rm bias}$, power injected to the VCSEL is represented by $\eta_2^{-1}(P_{\rm t})$, where $\eta_2^{-1}$ is the operation for the inverse function. Assume the efficiency of bias-Tee is $\eta_1$, power required at the transmitter including DC and signal power is expressed by
\begin{equation}
    P_{\rm T} = \frac{\eta_2^{-1}(P_{\rm t})}{\eta_1}.
\end{equation}

If the power consumed by  signal processing, analog-to-digital converter (ADC) and digital-to-analog converter (DAC) process are altogether denoted by $P_{\rm others}$, and DC power may be consumed at receiver side by TIA is denoted by $P_{\rm R}$, the consumption factor (CF) of VCSEL-based OWC system is
\begin{equation}
    {\rm CF_{\rm owc}} = \frac{B\log_2(1+\gamma_{\rm owc})}{P_{\rm T}+P_{\rm R}+P_{\rm others}}.
\end{equation}

\subsection{SISO link: TeraCom}
\subsubsection{Channel gain model}
We adopt a multi-ray channel model for TeraCom which utilizes the principles of
geometric optics to trace the propagation of line-of-sight (LoS), reflected,
diffusely scattered, and diffracted electromagnetic (EM) waves. Assume there are $N_{\rm Ref}$ reflected rays, $N_{\rm Sca}$ scattered rays, $N_{\rm Dif}$ diffracted rays, the transfer function of the channel at frequency $f$ is~\cite{multiRay}
\begin{equation}
    H(f) = H_{\rm LoS}(f)+\sum_{p=1}^{N_{\mathrm{Ref}}}H^{(p)}_{\rm Ref}(f)+\sum_{q=1}^{N_{\mathrm{Sca}}}H^{(q)}_{\rm Sca}(f)+\sum_{u=1}^{N_{\mathrm{Dif}}}H^{(u)}_{\rm Dif}(f),
\end{equation}
where $H_{\rm LoS}(f),H^{(p)}_{\rm Ref}(f),H^{(q)}_{\rm Sca}(f)$, and $H^{(u)}_{\rm Dif}(f)$ are transfer functions for the LoS, reflected, scattered and diffracted propagation paths, respectively, which will be specified below.

$H_{\rm LoS}$ relates to the antenna gain, spreading loss (SL), and molecular absorption loss (MAL), which is expressed by
\begin{equation}
\begin{aligned}
    H_{\rm LoS}(f) &= A_{\rm LoS} H_{\rm SL}(f)H_{\rm MAL}(f){\rm e}^{-j2\pi f\tau_{\rm LoS}}\\
    &=\sqrt{G_{\rm LoS}^{\rm Tx}G_{\rm LoS}^{\rm Rx}}\frac{c}{4\pi fr}{\rm e}^{\frac{1}{2}\kappa (f)r}{\rm e}^{-j2\pi f\tau_{\rm LoS}}
\end{aligned}
\label{eq:losthz}
\end{equation}
where $r$ is the LoS distance between transmitter and receiver and time delay $\tau_{\rm LoS}=r/c$. $\kappa (f)$ is a frequency-dependent medium absorption coefficient modeled specifically for $100-450$ GHz frequency band in~\cite{MAL}. $G_{\rm LoS}^{\rm Tx}$ and $G_{\rm LoS}^{\rm Rx}$ are angular-dependent antenna gain patterns at transmitter and receiver, respectively, which can be simplified by a Gaussian beam model as~\cite{antennapattern}
\begin{equation}
    G(\theta_{\rm a}, \theta_{\rm e})=G_0 \cdot {\rm e}^{-\left(\frac{\theta_{\rm a}-\theta_{\rm a}^0}{\theta_{\rm a}^{\rm bw}}\right)^2} \cdot {\rm e}^{-\left(\frac{\theta_{\rm e}-\theta_{\rm e}^0}{\theta_{\rm e}^{\rm bw}}\right)^2},
    \label{eq:antenna}
\end{equation}
where $\theta_{\rm a}^{\rm bw}$ and $\theta_{\rm e}^{\rm bw}$ describe the half-power beamwidth (HPBW) of the radiated pattern in the azimuth and elevation, $\theta_{\rm a}^0$ and $\theta_{\rm e}^0$ are the main beam direction. $\theta_{\rm a}, \theta_{\rm e}$ are the azimuth and elevation angles from the source point to the target point, and $G_0$ is the maximum gain. Eq.~\eqref{eq:antenna} is experimentally validated for horn antennas, where relation between $\theta_{\rm a,e}^{\rm bw}$ and $G_0$ is obtained approximately assuming a square antenna aperture as
\begin{equation}
    \theta_{\rm a,e}^{\rm bw} = 2\sqrt{\frac{\pi}{G_0}}.
\end{equation}

Transfer functions for reflected, scattered, and diffracted ray propagation are given as follows:
\begin{equation}
\begin{aligned}
        H_{\rm Ref}(f)&=\frac{A_{\rm Ref} R(f) c}{4\pi f (r_1+r_2)}{\rm e}^{-j2\pi \tau_{\rm Ref}-\frac{1}{2}\kappa (f)(r_1+r_2)}\\
        H_{\rm Sca}(f)&=\frac{A_{\rm Sca} S(f) c}{4\pi f (s_1+s_2)}{\rm e}^{-j2\pi \tau_{\rm Sca}-\frac{1}{2}\kappa (f)(s_1+s_2)}\\
        H_{\rm Dif}(f)&=\frac{A_{\rm Dif} L(f) c}{4\pi f (d_1+d_2)}{\rm e}^{-j2\pi \tau_{\rm Dif}-\frac{1}{2}\kappa (f)(d_1+d_2)}
\end{aligned},
\end{equation}
where $A_{\rm Ref}, A_{\rm Sca} , A_{\rm Dif}$ denote the antenna gains in reflected, scattered, and diffracted ray propagation, respectively. $\{r/s/d\}_{1}$ and $\{r/s/d\}_{2}$ represent the distance from the transmitter to the reflecting/scattering/diffracting point and the distance from these points to the receiver. $\tau_{\rm Ref} = \tau_{\rm LoS}+(r_1+r_2-r)/c$ and $\tau_{\rm Sca}, \tau_{\rm Dif}$ are calculated in the same manner. Besides, $R(f), S(f), L(f)$ denote reflection, scattering, and diffraction coefficient for rough surface, of which the details can be found in \cite{multiRay}. Given a room geometry and transmitter/receiver locations, reflected point locations are calculated. Then, scatterers are assumed as a set of points surrounding reflecting points. In indoor environments, diffraction effects are negligible as the region is close to the incident shadow boundary~\cite{multiRay}. Finally, the channel gain for TeraCom is $H_{\rm thz} = |H(f)|^2$.
\subsubsection{Signal-to-noise ratio} We take an electronics-based TeraCom system as an example, which generally has a symmetrical transceiver design. At the transmitter, the baseband signal is modulated to an intermediate frequency (IF) by the modulator at first, and then amplified by an IF power amplifier (PA) and mixed with the local oscillator (LO) signal by a mixer. After bandpass filter (BPF), the THz signal is transmitted by an antenna. At the receiver, captured by the antenna, the downconversion is applied to THz waves which is exactly the reverse process in the transmitter.
Assuming an additive white Gaussian noise (AWGN) channel, the received SNR at the receiver is defined as
\begin{equation}
    \gamma_0 = \frac{P_{\rm rec}}{\sigma_w^2}=\frac{P_{\rm rec}}{kTBF},
\end{equation}
where $P_{\rm rec}$ is the received signal power and $F$ is the noise figure at the receiver. However, the phase noise (PN) of local oscillators (LO) is negligible in TeraCom systems. To numerically calculate the PN effect after applying the existing OFDM PN mitigation schemes, a theoretical upper bound for the signal-to-interference-and-noise ratio (SINR) of the OFDM system affected by PN is adopted with $K_0$ indicating the PN floor in dBc/Hz as~\cite{PHN}
\begin{equation}
    \gamma_{\rm thz} = \frac{1}{2(1-{\rm e}^{-\frac{K_0B}{4}})+\gamma_0^{-1}}.
    \label{eq:thzsinr}
\end{equation}
\subsubsection{Consumption factor} As power flows in the typical electronics-based TeraCom system, components IF PA, BPF, and Mixer (M) consume power, which is represented by their power gain $G_{\rm \{PA, BPF, M\}}$ and efficiency $\eta_{\rm \{PA, BPF, M\}}$, where efficiency is defined as the ratio of the useful signal power to the total power consumption. Besides, DC power $P_{\rm DC}$ is required to drive LO at both transceivers. Thus, the CF of the TeraCom system is formulated as
\begin{equation}
    {\rm CF_{\rm thz}} = \frac{B\log_2(1+\gamma_{\rm thz})}{\frac{P_{\rm rec}}{H_{\rm link}}+2P_{\rm DC}+P_{\rm others}}.
\end{equation}
$H_{\rm link}$ is then given by the combination of the transceivers' power efficiency $H_{\rm Tx}$ and  $H_{\rm Rx}$ as~\cite{consumption}
\begin{equation}
    H_{\rm link}^{-1} = H_{\rm Rx}^{-1}+\frac{1}{G_{\rm rec}}\left(\frac{1}{H_{\rm thz}}-1\right)+\frac{1}{G_{\rm rec}H_{\rm thz}}\left(H_{\rm Tx}^{-1}-1\right),
\end{equation}
where $G_{\rm rec}$ is the receiver gain, and $H_{\rm Tx}$ is expressed as
\begin{equation}
    H_{\rm Tx} = \left\{ 1+ \sum_{n=1}^{N}\frac{1}{\prod\limits_{i=n+1}^{N}G_i}\left(\frac{1}{\eta_n}-1\right)\right\}^{-1},
\end{equation}
where $N$ is the number of cascaded components in the transmitter and in this case $N=4$. $i,n\in \{1,2,3,4\}$ are indices of gain and efficiency vectors $\left\{ G_{\rm PA}, G_{\rm BPF1},G_{\rm M},G_{\rm BPF2}\right\}$ and $\left\{ \eta_{\rm PA}, \eta_{\rm BPF1},\eta_{\rm M},\eta_{\rm BPF2}\right\}$. $H_{\rm Rx}$ is defined in a same manner.

\subsection{Networked systems}
As mentioned above, the high directivity elements should be embedded at both TeraCom and VCSEL-based OWC systems to compensate for the severe path loss and obtain better communication performance with energy-concentrated beam transmission. Besides utilizing advanced beam steering techniques which are still under development for both technologies, multiple access points (APs) are required in a given room to guarantee service coverage. Hence, we derive analytical models for coverage probability and energy efficiency in networked VCSEL-based OWC and TeraCom systems. 
\subsubsection{Distributed AP design} The distributed AP deployment scheme is considered here: APs are distributed normally on the room's ceiling, where the beam direction of each antenna or each VCSEL transmitter is towards the floor, while the receiver's location is random. For user connection, the nearest AP user association is utilized, which means that the user selects the nearest
AP to associate with and the others are interfering APs.

\subsubsection{Coverage probability}
In networked systems, noise generated by interfering APs should be considered in the SINR models. Suppose there are $N_{\rm t}\times N_{\rm t}$ APs and $N_{\rm r}$ receivers, the $l$-th AP and $m$-th receiver form a transmission link $H_{\rm owc}^{(lm)}$ or $H_{\rm thz}^{(lm)}$. For networked-based VCSEL-based OWC systems, Eq.~\eqref{eq:owcsnr} is modified to SINR of the link between the tagged receiver and its associated AP $H_{\rm owc}^{(l'm')}$ as
\begin{equation}
    \gamma_{\rm owc}^{(m')} = \frac{R_{\rm PD}^2(H_{\rm owc}^{(l'm')})^2P_{\rm sig}}{\sum_{l\neq l'}^{N_{\rm{t}}\times N_{\rm{t}}} R_{\rm PD}^2(H_{\rm owc}^{(lm')})^2P_{\rm sig} + (\sigma^{m'})^2},
\end{equation}
where noise variance is also modified as
\begin{equation}
    \begin{aligned}
        (\sigma^{m'})^2 = &\frac{4kTBF_{\rm n}}{R_{\rm L}}+2qB\left(\sum_{l=1}^{N_{\rm{t}}\times N_{\rm{t}}}R_{\rm PD}H_{\rm owc}^{(lm')}P_{\rm t}\right)\\&+BN_{\rm RI}\left(\sum_{l=1}^{N_{\rm{t}}\times N_{\rm{t}}}(R_{\rm PD}H_{\rm owc}^{(lm')}P_{\rm t})^2\right).
    \end{aligned}
\end{equation}
Then, the coverage probability (CP) of the networked VCSEL-based OWC system is given by
\begin{equation}
    {\rm CP}_{\rm owc}(\gamma_{\rm th}) = \mathbb{P} (\gamma_{\rm owc}^{(m')} > \gamma_{\rm th}),
    \label{eq:cpvcsel}
\end{equation}
where $\gamma_{\rm th}$ indicates threshold SINR and $\mathbb{P}(\cdot)$ is the probability density function.

Similarly, interference noises should also be added to the SINR model of TeraCom systems. According to Eq.~\eqref{eq:thzsinr}, SINR of $m'$-th receiver is rewritten as
\begin{equation}
    \gamma_{\rm thz}^{(m')} = \left[{2(1-{\rm e}^{-\frac{K_0B}{4}})+\frac{\sum\limits_{l\neq l'}^{N_{\rm{t}}\times N_{\rm{t}}} P_{\rm rec}^{(lm')}+\sigma_w^2}{P_{\rm rec}^{(l'm')}}}\right]^{-1},
\end{equation}
where $P_{\rm rec} =P_{\rm t}H_{\rm thz}^{(lm')}\prod_{i=1}^{N} G_i$ is related to the transmitted power from TeraCom transmitter $P_{\rm t}$, channel gain, and gains of components at the receiver. ${\rm CP}_{\rm thz}(\gamma_{\rm th})$ is then calculated in the same manner as Eq.~\eqref{eq:cpvcsel}.
\subsubsection{Energy efficiency} To evaluate the energy efficiency of the networked system, CF of VCSEL-based OWC is given by
\begin{equation}
    {\rm CF}_{\rm owc} = \frac{\sum_{m=1}^{N_{\rm r}} B\log_2(1+\gamma_{\rm owc}^{(m)})}{\sum_{l=1}^{N_{\rm t}^2}  (P_{\rm T}^{(l)}+P_{\rm others}^{(l)}) + \sum_{m=1}^{N_{\rm r}}P_{\rm R}^{(m)}}.
\end{equation}
For TeraCom networked system, the CF is expressed as
\begin{equation}
    {\rm CF}_{\rm thz} = \frac{\sum_{m=1}^{N_{\rm r}} B\log_2(1+\gamma_{\rm thz}^{(m)})}{\sum\limits_{l=1}^{N_{\rm t}^2}  \left(\frac{P_{\rm t}^{(l)}}{H_{\rm Tx}}+P_{\rm DC}^{(l)}+P_{\rm others}^{(l)}\right) + \sum\limits_{m=1}^{N_{\rm r}}P_{\rm DC}^{(m)}}.
\end{equation}

\begin{table*}[ht]
\centering
\caption{Parameters for evaluation in indoor scenarios\label{long}}
\resizebox{\textwidth}{38mm}{
\begin{tabular}{ccc|ccc}
\hline
\multicolumn{3}{c|}{TeraCom} & \multicolumn{3}{c}{VCSEL-based OWC} \\ \hline
Parameter      & Symbol      & Value     & Parameter      & Symbol      & Value      \\ \hline
Frequency  &  $f$   &    $350$GHz    &Wavelength  & $\lambda$    &   $940$nm     \\ 
-  &  -    &   -   &  Original beam waist &  $\omega_{0}$    &  $2.523 \mu$m\\
PN floor  &  $K_0$    &   $-110$dBc/Hz   &  Relative intensity noise       &  $N_{\rm RI}$   &   $-155$dB/Hz\\
Rx noise figure  &  $F$    &   $10.6$dB   &  TIA noise figure          &  $F_{\rm n}$    &   $5$dB\\
Gain of PA    &$G_{\rm PA}$  &    $10.9$dB   &  PD responsivity          &  $R_{\rm PD}$    &   $0.6$A/W\\ 
Power efficiency of PA    &$\eta_{\rm PA}$  &   $0.1165$   &  Load resistance          &  $R_{\rm L}$    &   $50 \Omega$\\ 
Gain/efficiency of mixer    &$G_{\rm M}/\eta_{\rm M}$  &   $-13$dB   &       Efficiency of bias tee    &  $\eta_1$     &  $-1$dB\\ 
Gain/efficiency of first BPF    &$G_{\rm BPF1}/\eta_{\rm BPF1}$  &   $-5$dB   &   VCSEL saturation power        &  $P_{\rm sat}$     &  $14.6$ mW\\ 
 Gain/efficiency of second BPF    &$G_{\rm BPF2}/\eta_{\rm BPF2}$  &  $-12.84$dB   &  VCSEL threshold current         &  $I_{\rm th}$    &  $2$ mA\\
 Power supply for LO    &$ P_{\rm DC}$  &  $100$mW  &  VCSEL input bias voltage         &  $V_{\rm bias}$    &  $2.7$ V\\
 Gain of receiver    & $G_{\rm rec}$  &  $0$dB  &  VCSEL conversion efficiency         &  $\eta$    &  $0.66$ \\
  -    &-  &  -  & Fitness of VCSEL conversion curve   &  $N_{\gamma}$    &  $2$ \\
  \hline
\end{tabular}}
\label{t:paraindoor}
\end{table*}
\subsection{Comparison based on numerical analysis}
For performance comparison of TeraCom and VCSEL-based OWC systems in indoor applications, a room with geometry of $3$m$\times 3$m$\times 3$m is configured. The transmitter (Tx) is mounted on the ceiling with a height of $2.95$ m and the receiver (Rx) is assumed to be placed at a height of $0.95$ m as the user equipment is generally on top of tables, etc. Hence, without specifying, the location of Tx is $[1.5, 1.5, 2.95]$ and Rx is $[1.5, 1.5, 0.95]$, respectively, so the link length of both systems is $2$m. Also, the Tx is facing downward and Rx vice versa. For SISO links, the horn antenna is employed in the system, and the HPBW of the radiated pattern in the azimuth and elevation are assumed to be identical, i.e., $\theta_{\rm a}^{\rm bw}=\theta_{\rm e}^{\rm bw}$ at both Tx and Rx sides. Besides, considering the orientation of Tx and Rx, $\theta_{\rm a}^{0}=0^{\circ}$ and $\theta_{\rm e}^{0}=90^{\circ}$ are set at the Tx side while  $\theta_{\rm a}^{0}=0^{\circ}$ and $\theta_{\rm e}^{0}=-90^{\circ}$ are set at the Rx side. 

To obtain the channel gain of the TeraCom system, multi-ray channel simulating follows the logic in \cite{multiRay}, where the key parameters for calculating $R(f), S(f), L(f)$ can be found. Additionally, we include the double-bounce reflection and scattering effect, neglecting the diffraction effect in indoor scenarios at the $350$GHz band. Parameters for evaluating TeraCom's SNR and CF, i.e., noises, gain/efficiency of key components, in the transceiver architecture are listed in Table.~\ref{t:paraindoor}. To numerically analyze the performance of VCSEL-based OWC, receiver-related parameters are from~\cite{VCSELtrans}, and the parameters for the VCSEL conversion process are obtained through the fitting process of measurement results~\cite{citatkazemiexperimental}. The parameters for the two systems used in the numerical analysis are summarized in Table.~\ref{t:paraindoor} and the commonly adopted parameters include: temperature in Kelvin $K=295$, power consumed by the signal processing components $P_{\rm others}=100$mW.
In the following, we will first compare the two systems by SNR and CF for single-input-single-output (SISO) links in a room. Then, we will analyze coverage probability as well as CF of the two systems when multiple Txs are deployed.
\subsubsection{SISO SNR}

\begin{figure*}  
\centering
\subfigure[THz, HPBW = $8^{\circ}$]{
\begin{minipage}[b]{0.3\textwidth}
\includegraphics[width=1\textwidth]{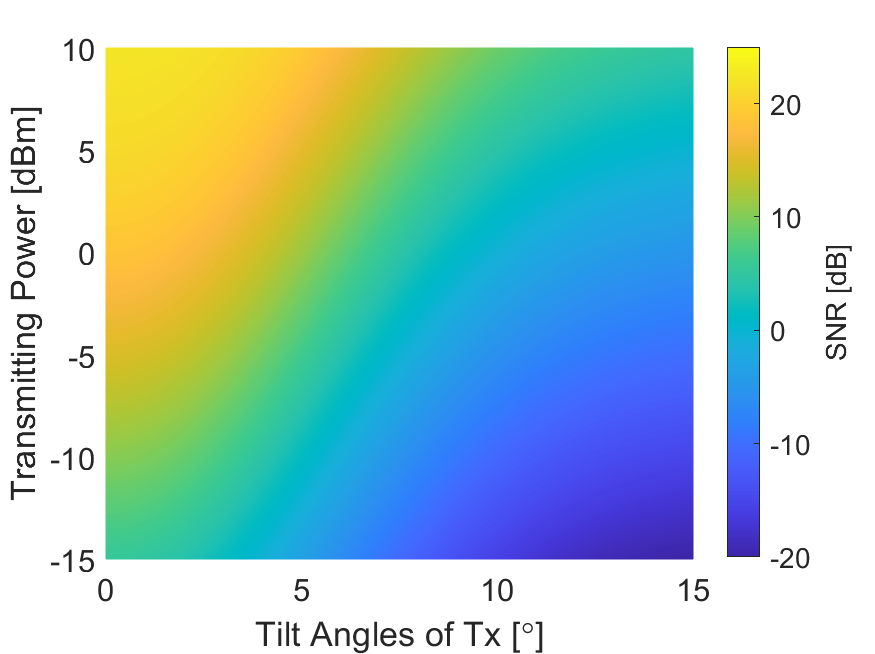}
\end{minipage}
}
\subfigure[THz, HPBW = $6^{\circ}$]{
\begin{minipage}[b]{0.3\textwidth}
\includegraphics[width=1\textwidth]{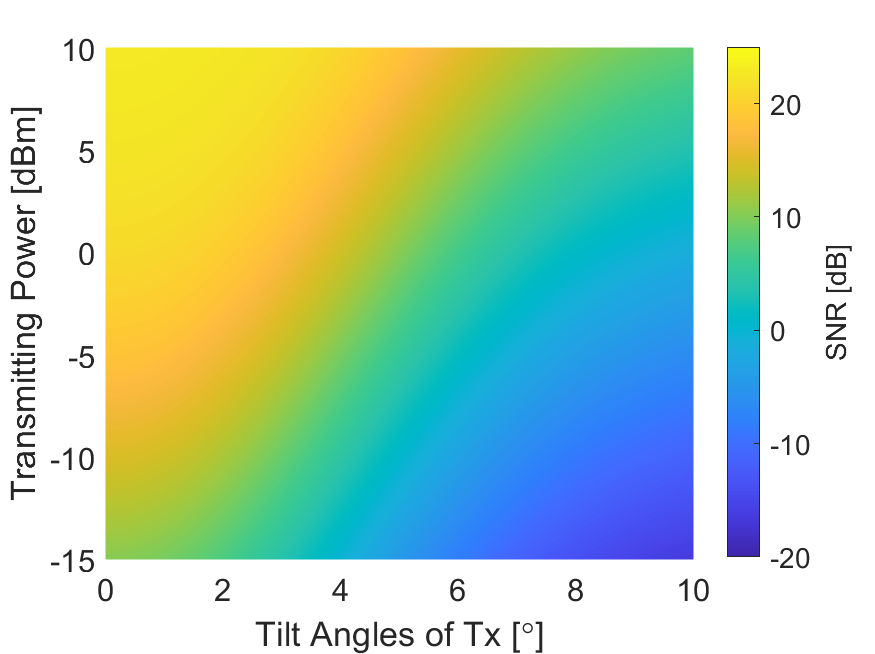}
\end{minipage}
}
\subfigure[THz, HPBW = $4^{\circ}$]{
\begin{minipage}[b]{0.3\textwidth}
\includegraphics[width=1\textwidth]{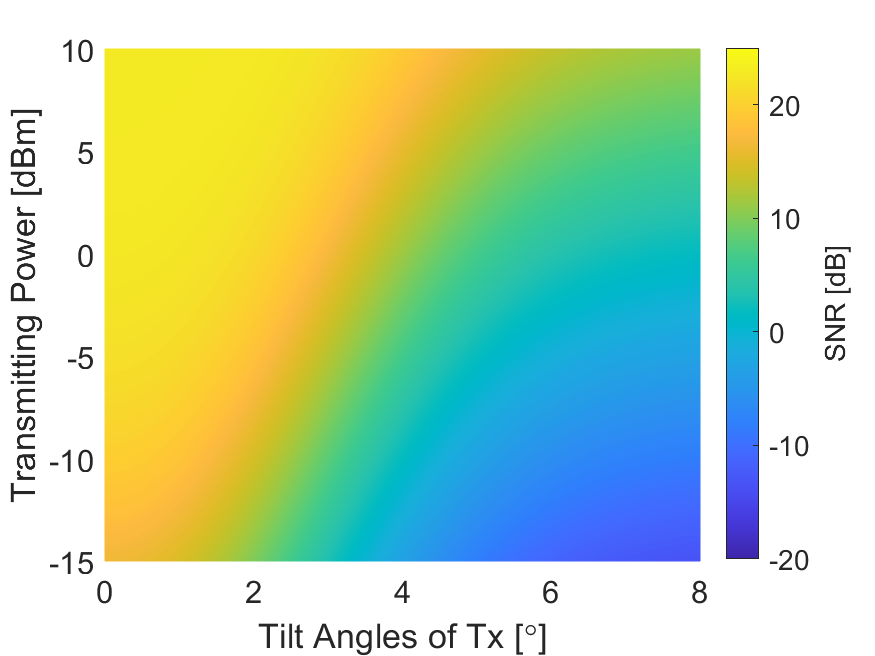}
\end{minipage}
}\\
\subfigure[VCSEL, $2\theta_{\rm CPC}$ = HPBD = $8^{\circ}$]{
\begin{minipage}[b]{0.3\textwidth}
\includegraphics[width=1\textwidth]{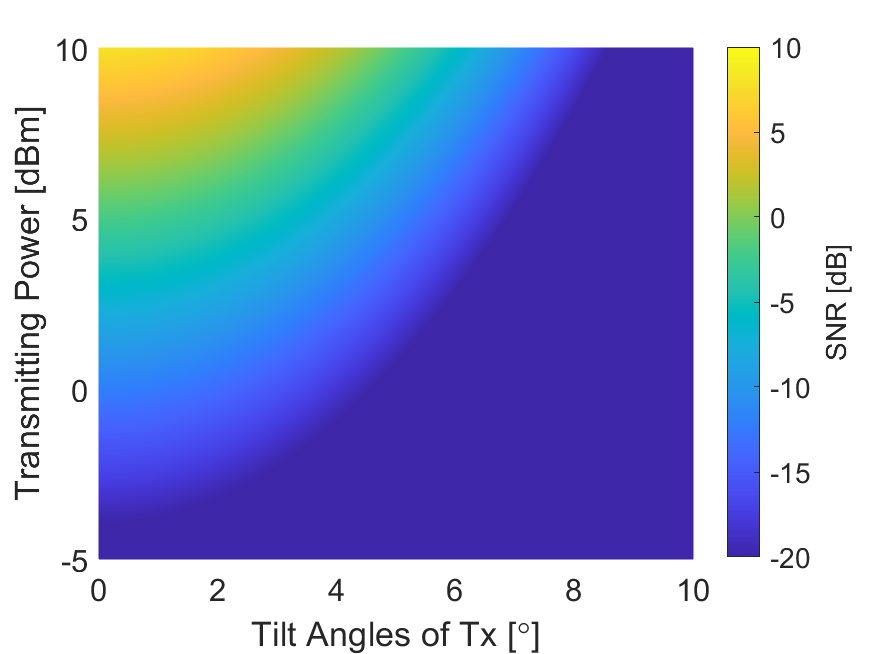}
\end{minipage}
}
\subfigure[VCSEL, $2\theta_{\rm CPC}$ = HPBD = $6^{\circ}$]{
\begin{minipage}[b]{0.3\textwidth}
\includegraphics[width=1\textwidth]{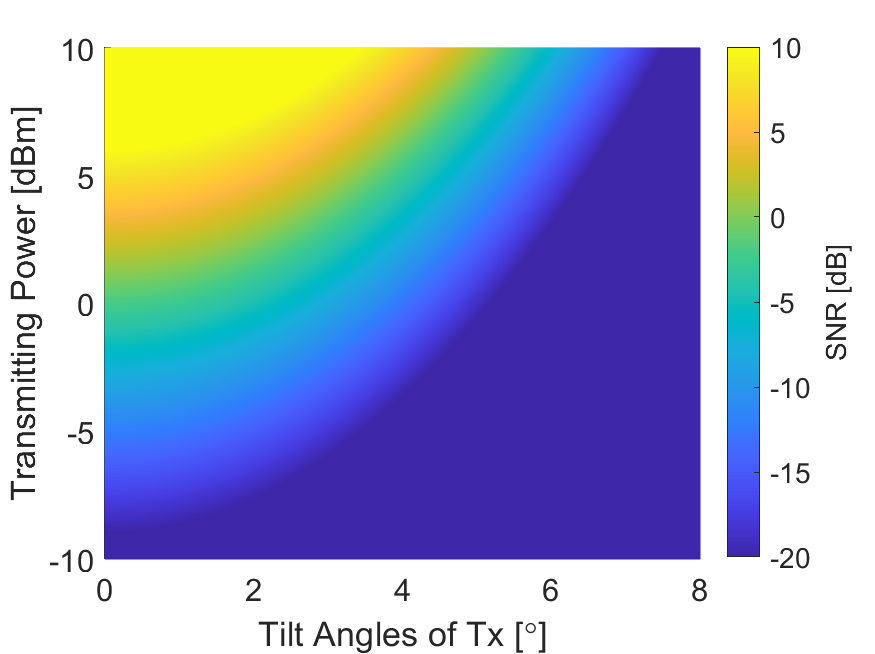}
\end{minipage}
}
\subfigure[VCSEL, $2\theta_{\rm CPC}$ = HPBD = $4^{\circ}$]{
\begin{minipage}[b]{0.3\textwidth}
\includegraphics[width=1\textwidth]{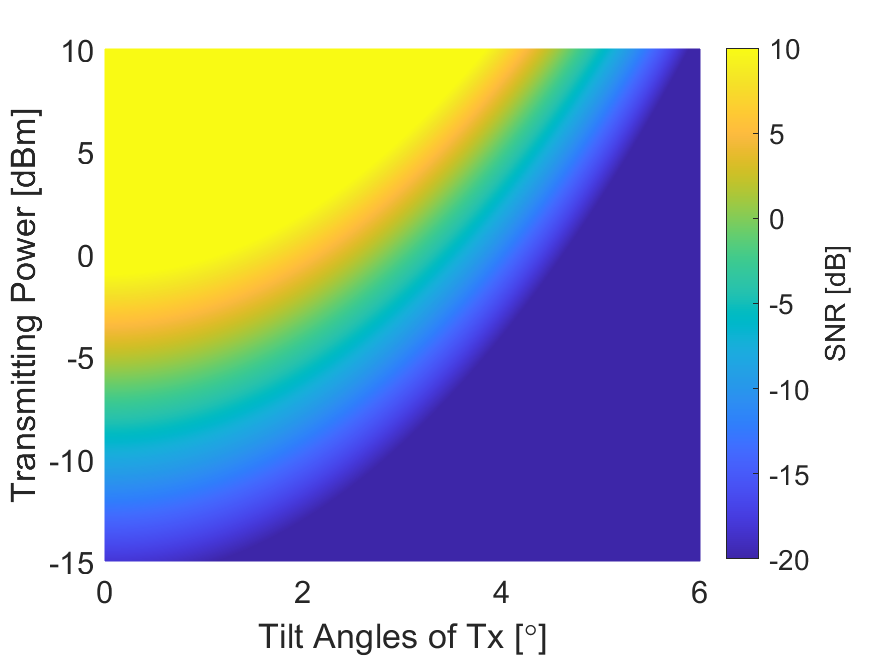}
\end{minipage}
}
\caption{SNR comparison of TeraCom and VCSEL-based OWC systems with various transmitting power and tilt angles of Tx, while the HPBW of TeraCom and HPBD of VCSEL-based OWC are the same as $4^{\circ}$, $6^{\circ}$, and $8^{\circ}$.}
\label{f:comptiltSNR}
\end{figure*}
To evaluate the SNR of the two systems specifically with misalignment effect, two parameters matter: channel gain as the Tx/Rx is tilted or displaced, and transmitted power. As we found the similarity in the tradeoff between beam coverage and gain of THz antenna and optical lens, we try to guarantee the same beam coverage for the two systems to make a fair comparison in the following numerical analysis, i.e., HPBW of THz antennas at both Tx and Rx sides, HPBD of VCSEL beams, and FoV of the Rx in VCSEL-based OWC system are the same. Moreover, the bandwidth $B=1$GHz is set for both systems. 

As in Fig.~\ref{f:comptiltSNR}, we evaluate the SNR of TeraCom and VCSEL-based OWC systems with various transmitting power $P_{\rm t}$ ranging from $-15$dBm to $10$dBm and tilted angles of Tx ranging from $0^{\circ}$ to $15^{\circ}$, while HPBW = HPBD = $2\theta_{\rm CPC} = 8^{\circ}, 6^{\circ}, 4^{\circ}$. From Figs.~\ref{f:comptiltSNR} (a)-(c), we can find that lower HPBW which means a more focused beam transmitted leads to higher SNR with lower transmitting power; for example, as HPBW=$8^{\circ}$, more than $-10$dBm transmitting power is required to obtain $>10$dB SNR if the transceivers are strictly aligned with each other, while no more than $-15$dBm is needed as HPBW=$4^{\circ}$. At the same time, it is evident that the system's tolerance to misalignment effects has decreased. If we also take $10$dB SNR as a benchmark, the maximum allowed tilt angles with $10$dBm transmitting power degrade from around $8^{\circ}$ to $4^{\circ}$. As in Figs.~\ref{f:comptiltSNR} (d)-(f), we can find similar trends in required transmitting power and misalignment tolerance for VCSEL-based OWC systems. However, SNR obtained with the same configuration but a larger HPBD is comparably lower and the required power is higher. However, with a more focused VCSEL beam, i.e., HPBD=$4^{\circ}$, the SNR remains consistently high throughout the system coverage area. From the comparison, two reasons leading to more required power in VCSEL-based OWC: i) for OWC, the optical beam intensity needs to be converted into electrical current through photodetector devices, which leads to the power of channel gain when obtaining SNR in VCSEL-based OWC system as in Eq.~\eqref{eq:owcsnr}; ii) for OWC, non-constraint requirement need to be satisfied so that the signal power generally lower than total transmitted beam power, i.e., relation between transmitting power and signal power is $P_{\rm sig}=\frac{1}{9}P_{\rm t}^2$ if OFDM is employed. 

\begin{figure*}  
\centering
\subfigure[THz, HPBW = $8^{\circ}$]{
\begin{minipage}[b]{0.3\textwidth}
\includegraphics[width=1\textwidth]{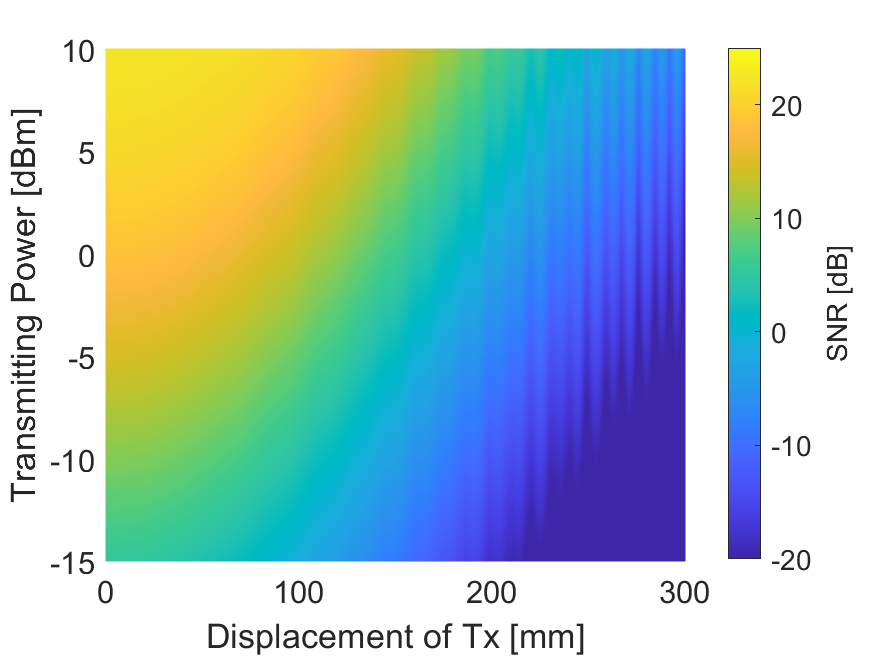}
\end{minipage}
}
\subfigure[THz, HPBW = $6^{\circ}$]{
\begin{minipage}[b]{0.3\textwidth}
\includegraphics[width=1\textwidth]{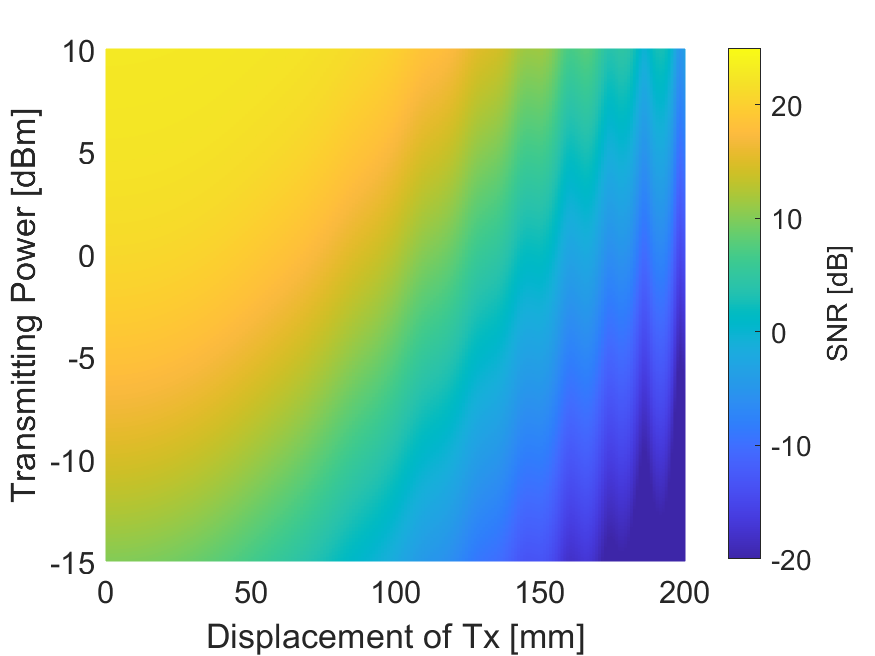}
\end{minipage}
}
\subfigure[THz, HPBW = $4^{\circ}$]{
\begin{minipage}[b]{0.3\textwidth}
\includegraphics[width=1\textwidth]{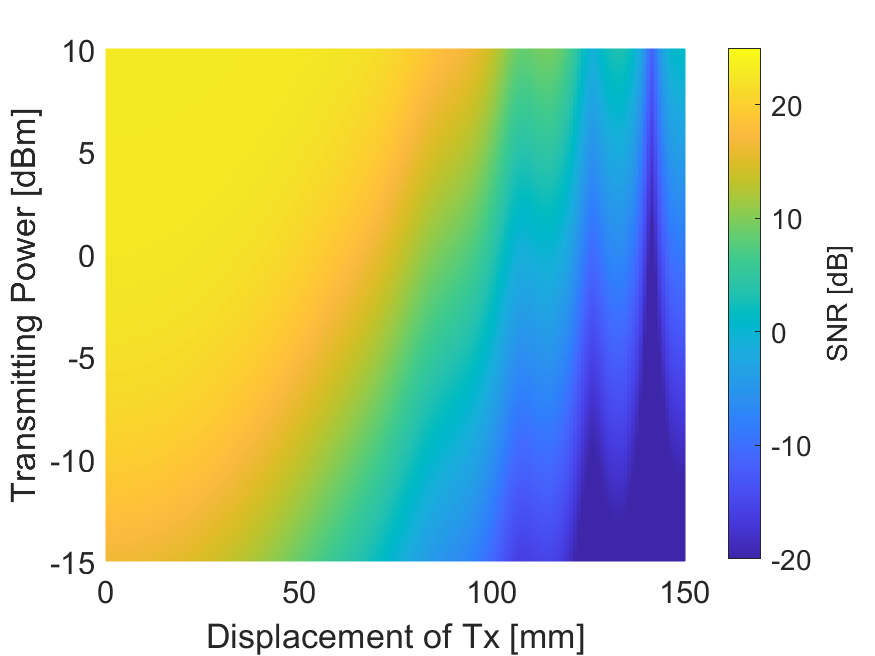}
\end{minipage}
}\\
\subfigure[VCSEL, $2\theta_{\rm CPC}$ = HPBD = $8^{\circ}$]{
\begin{minipage}[b]{0.3\textwidth}
\includegraphics[width=1\textwidth]{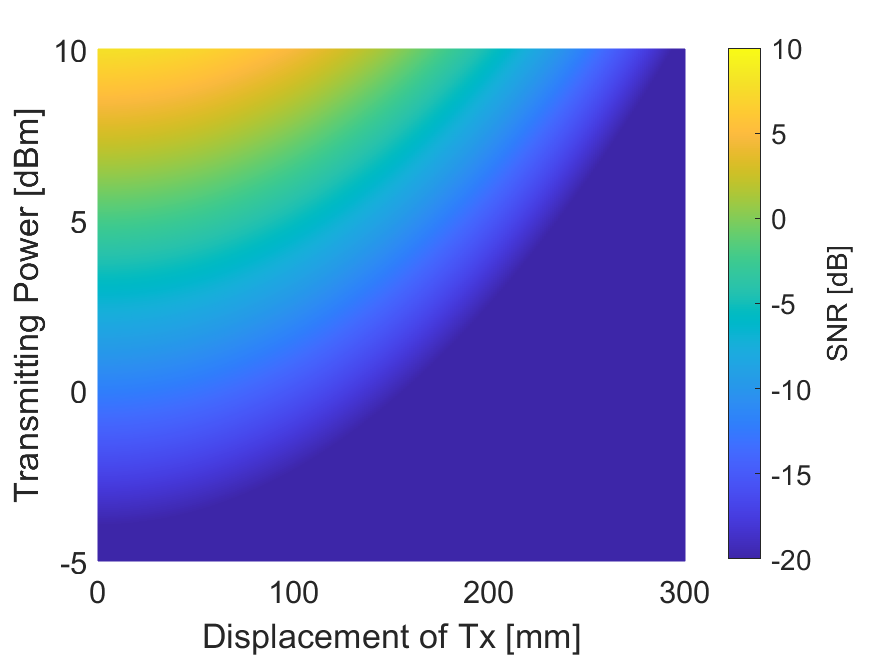}
\end{minipage}
}
\subfigure[VCSEL, $2\theta_{\rm CPC}$ = HPBD = $6^{\circ}$]{
\begin{minipage}[b]{0.3\textwidth}
\includegraphics[width=1\textwidth]{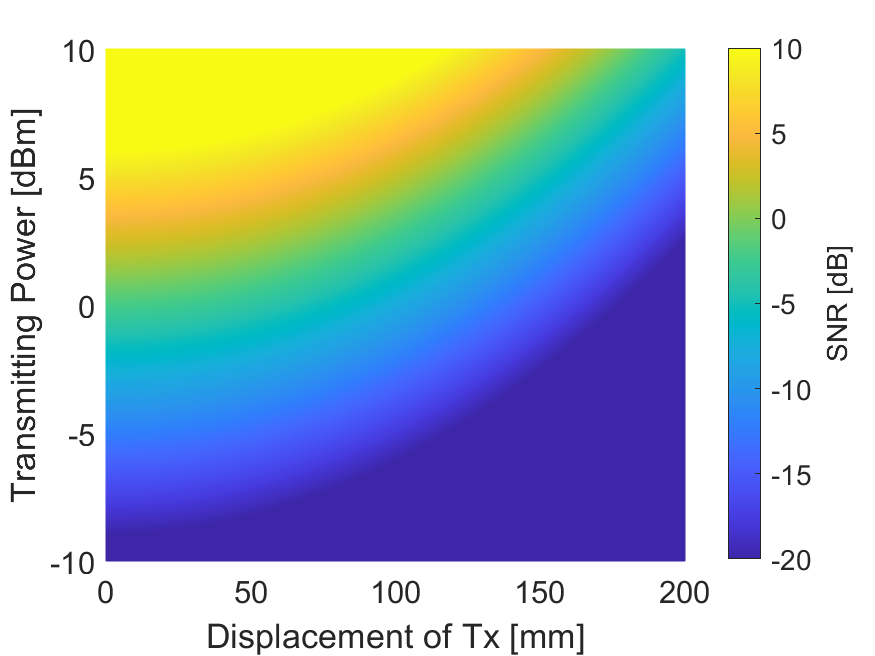}
\end{minipage}
}
\subfigure[VCSEL, $2\theta_{\rm CPC}$ = HPBD = $4^{\circ}$]{
\begin{minipage}[b]{0.3\textwidth}
\includegraphics[width=1\textwidth]{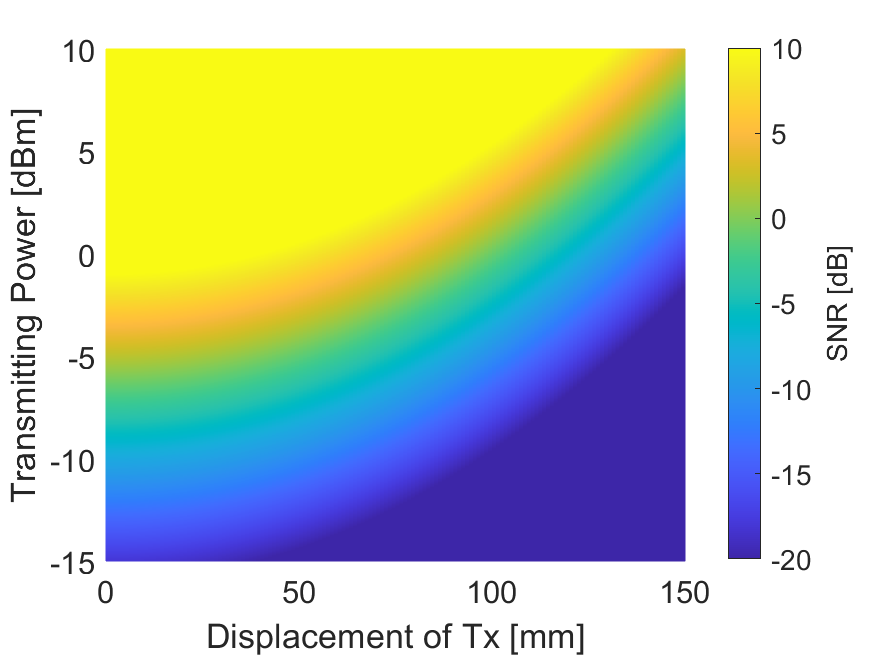}
\end{minipage}
}
\caption{SNR comparison of TeraCom and VCSEL-based OWC systems with various transmitting power and displacement distance along x-axis of Tx, while the HPBW of TeraCom and HPBD of VCSEL-based OWC are the same as $4^{\circ}$, $6^{\circ}$, and $8^{\circ}$.}
\label{f:compdisSNR}
\end{figure*}

The same simulations are conducted for the comparison while Tx is displaced with a distance along the x-axis ranging from $0-400$mm as in Fig.~\ref{f:compdisSNR}. We can find similar trends as in the case where the Tx is with tilt angles. The SNR results of VCSEL-based OWC systems, as shown in Figs.~\ref{f:compdisSNR} (d)-(f), can be correlated to those with a tilted transmitter through geometric relationships. However, as in Figs.~\ref{f:compdisSNR} (a)-(c), the TeraCom system appears to have a weaker tolerance to misalignment. This is because the displacement of the Tx is equivalent to both the Tx and Rx antennas having a tilt angle. Besides, we can see stripes in Figs.~\ref{f:compdisSNR} (a)-(c), which are caused by the multi-path effect of THz wave propagation.

Although we plot the SNR for both systems with varying transmitting power, the maximum achievable and allowable transmitting power of the TeraCom and VCSEL-based OWC systems represent their respective limitations. For the TeraCom system, the achievable transmitting power from the transmitter is still limited at high frequencies. As can be found in the literature, the transmitting power at $350$GHz is still around $40\mu$W. However, for VCSEL-based OWC systems, laser beams can be generated with super high power while the eye safety regulations set the limits for the maximum allowed transmitting power. Considering the most conservative requirement with the original beam waist and optical lenses, the transmitting power is set at $P_{\rm t}=1$mW.

\begin{figure}
    \centering
    \includegraphics[width=0.45\linewidth]{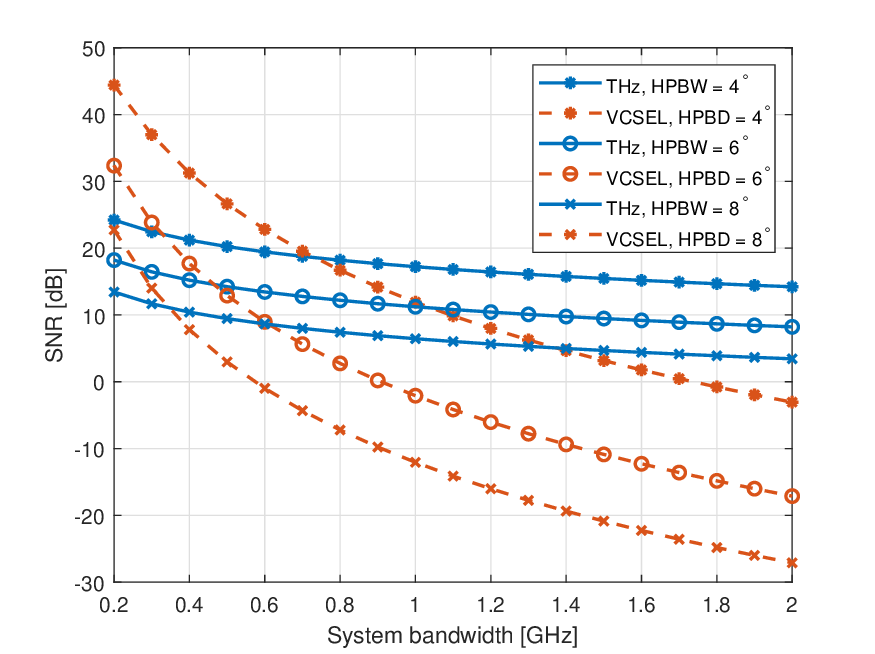}
    \caption{SNR comparison of TeraCom and VCSEL-based OWC systems with various system bandwidth, while the HPBW of TeraCom and HPBD of VCSEL-based OWC are the same as $4^{\circ}$, $6^{\circ}$, and $8^{\circ}$.}
    \label{f:snrbandwidth}
\end{figure}
As bandwidth is also a limitation of VCSEL-based OWC systems, we compare the SNR of the two systems with varying bandwidth as depicted in Fig.~\ref{f:snrbandwidth}, which in VCSEL-based OWC systems is mainly restricted by receiving PD and in TeraCom systems impacts system noises. In this simulation, the transmitting power for the VCSEL beam is set at $P_{\rm t}^{\rm v} = 1$mW, while the power for transmitting THz waves is $P_{\rm t}^{\rm thz}=40\mu$W. The results indicate a decline in SNR with increasing system bandwidth in both systems, with the VCSEL-based OWC system being more sensitive to changes in bandwidth. Additionally, we can see that reducing the HPBD by using optical lenses to create a more focused beam can effectively address this issue, allowing for higher bandwidth while maintaining the required SNR. Still, it's challenging for VCSEL-based OWC to outperform the TeraCom systems if the bandwidth exceeds 1 GHz with the current receiver design. Meanwhile, numerous studies, demonstrations, and products focused on PD design with higher bandwidth have emerged to address this issue.


\subsubsection{SISO CF}
\begin{figure*}  
\centering
\subfigure[$B=500$ MHz, $L=2$ m]{
\begin{minipage}[b]{0.3\textwidth}
\includegraphics[width=1\textwidth]{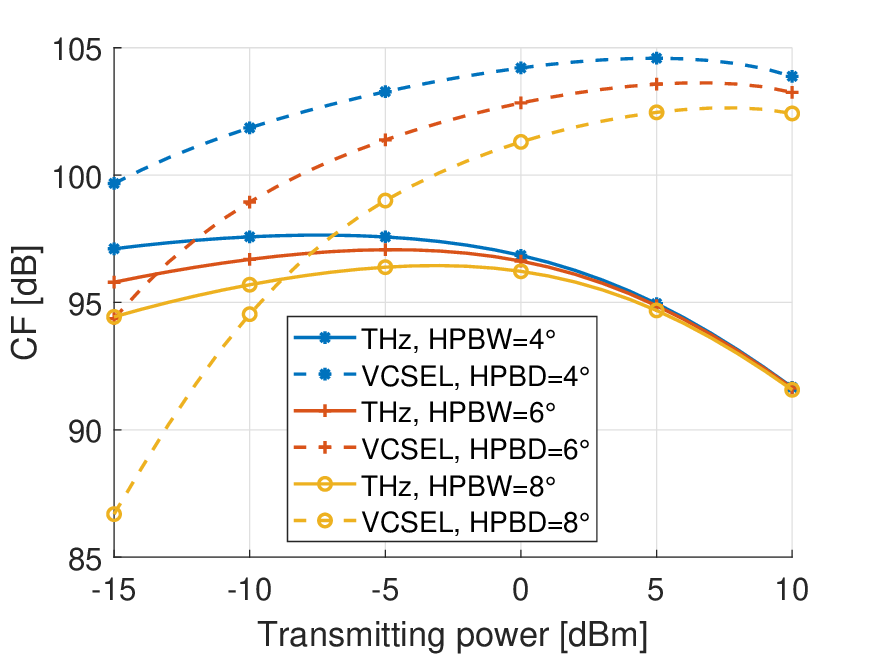}
\end{minipage}
}
\subfigure[$L=2$ m]{
\begin{minipage}[b]{0.3\textwidth}
\includegraphics[width=1\textwidth]{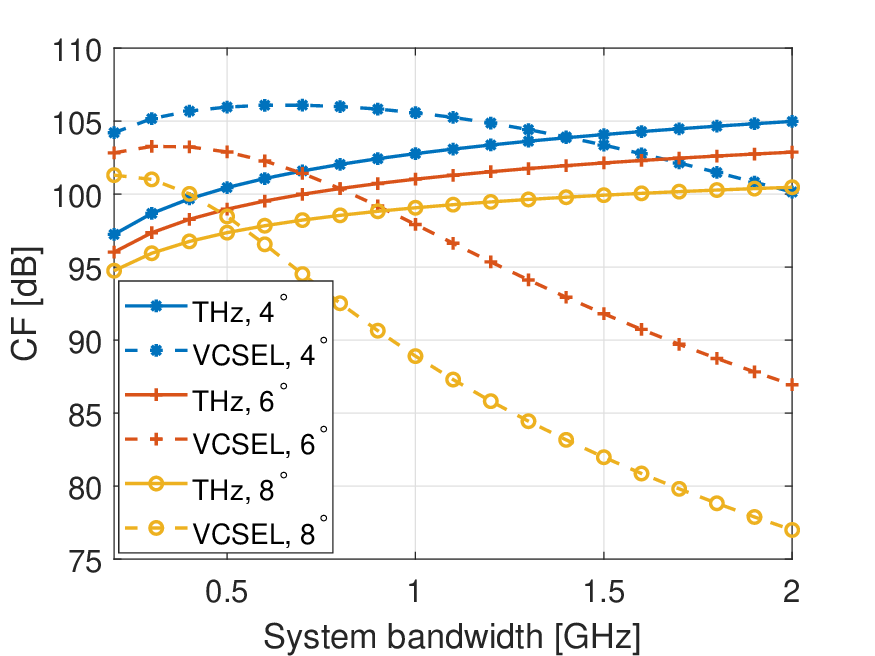}
\end{minipage}
}
\subfigure[$B=500$ MHz]{
\begin{minipage}[b]{0.3\textwidth}
\includegraphics[width=1\textwidth]{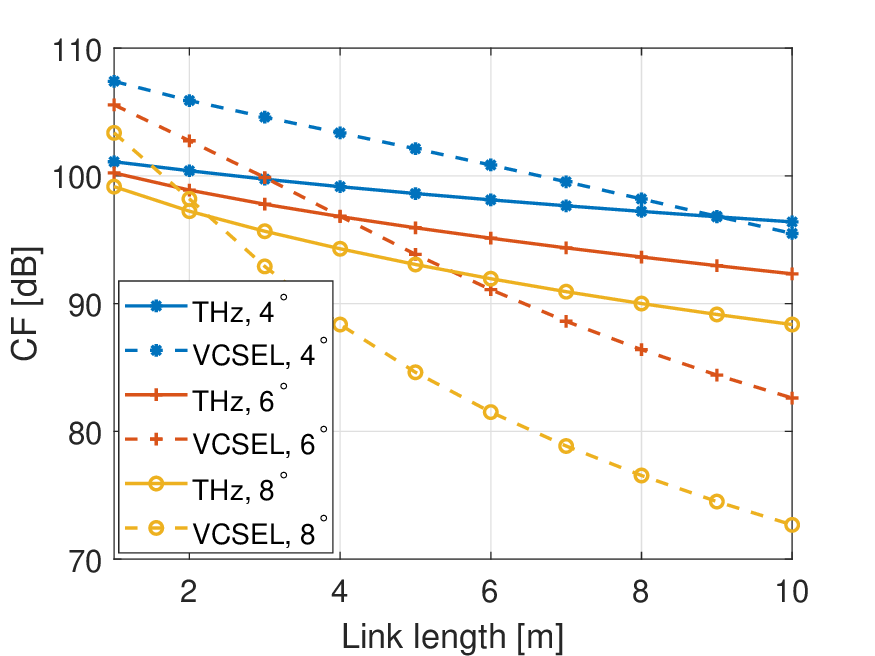}
\end{minipage}
}
\caption{CF comparison of TeraCom and VCSEL-based OWC systems under varying transmitting power, system bandwidth, and link length.}
\label{f:cfcompaligned}
\end{figure*}

Energy efficiency is a crucial metric for next-generation networks. First, we compare the CF of the two systems under varying transmission power, system bandwidth, and link length to analyze the factors influencing system performance assuming Tx and Rx are strictly aligned with each other, as in Fig.~\ref{f:cfcompaligned}. Still, we standardize the HPBW of antenna patterns at both the Tx and Rx sides in the TeraCom system, the HPBD of VCSEL beams, and the FoV of VCSEL-based OWC receivers to evaluate system performance under the same conditions for handling misalignment. Fig.~\ref{f:cfcompaligned} (a) depicts the CF of the two systems as a function of transmitting power. As the beam-focusing ability, i.e., beam energy concentration, increases, the channel gain improves, enhancing CF performance for both systems. However, beam energy concentration matters more in VCSEL-based OWC systems. Higher beam divergence requires higher transmitting power to ensure the system's performance.
Additionally, as the transmitting power increases, the CF for both systems initially rises and then declines. In VCSEL-based OWC systems, the saturation of electro-optic conversion in the VCSEL transmitter contributes to this decline. In contrast, in TeraCom systems, the dominance of DC power consumed by cascaded components at the transceivers leads to a decrease. 
\begin{figure*}  
\centering
\subfigure[Tilt angle $0^{\circ}$]{
\begin{minipage}[b]{0.3\textwidth}
\includegraphics[width=1\textwidth]{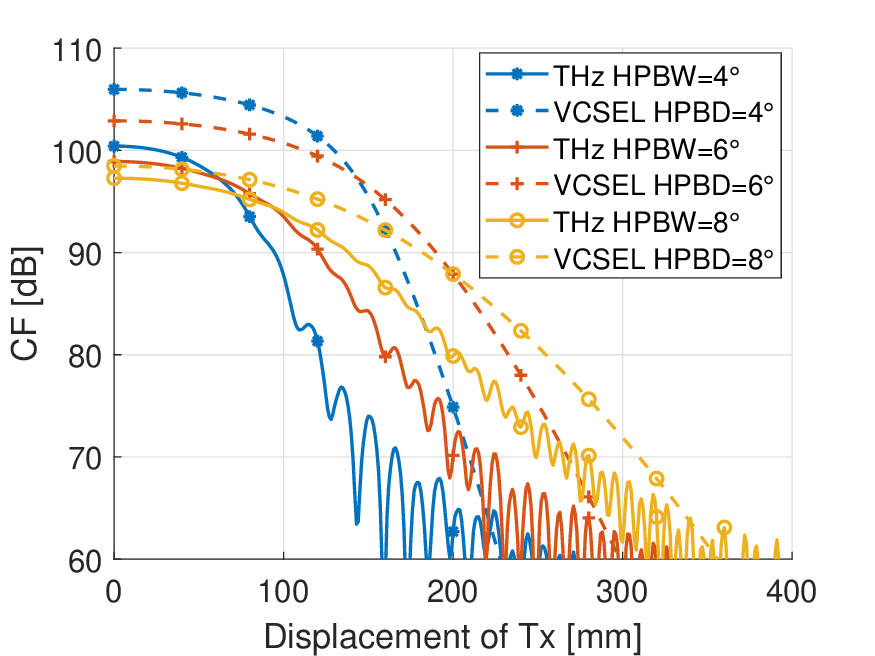}
\end{minipage}
}
\subfigure[Displacement $0$ mm]{
\begin{minipage}[b]{0.3\textwidth}
\includegraphics[width=1\textwidth]{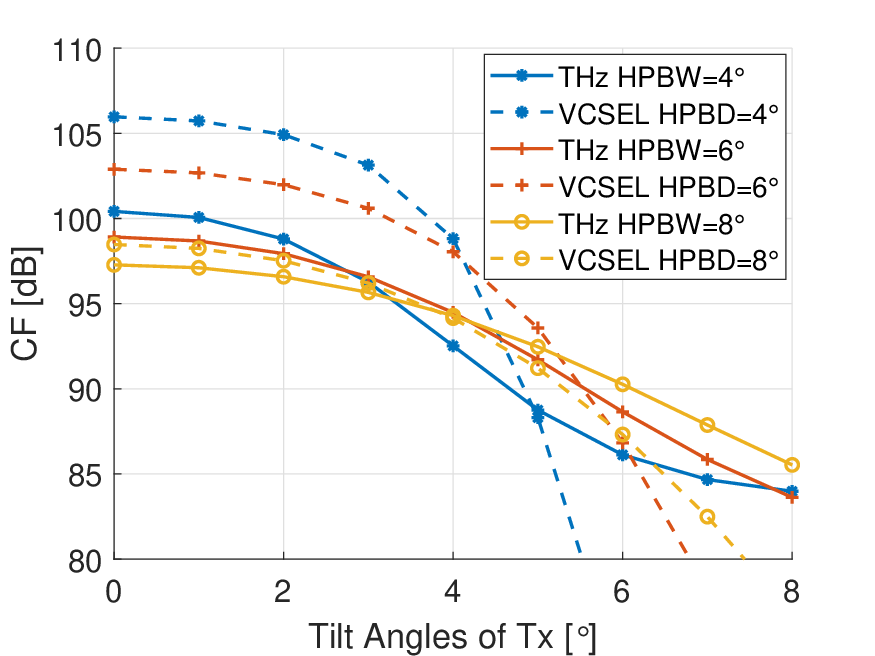}
\end{minipage}
}
\subfigure[Tilt angle $2^{\circ}$]{
\begin{minipage}[b]{0.3\textwidth}
\includegraphics[width=1\textwidth]{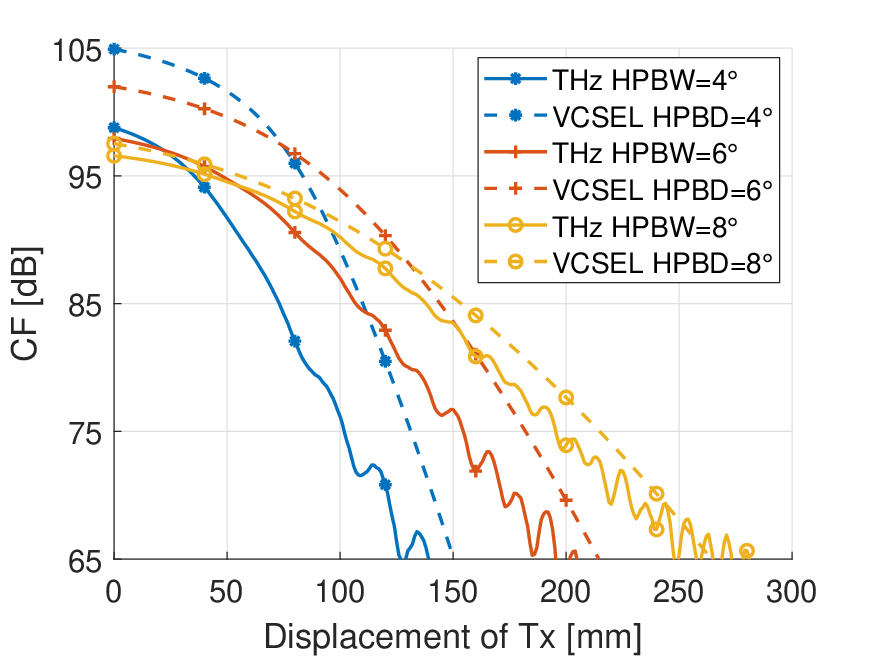}
\end{minipage}
}
\caption{CF comparison of TeraCom and VCSEL-based OWC systems with misalignment between Tx and Rx.}
\label{f:cfcompmisaligned}
\end{figure*}
We use the maximum achievable and allowable transmitting power for both systems to analyze the impacts of bandwidth and link length, respectively. In Fig.~\ref{f:cfcompaligned} (a), the CF of the VCSEL-based OWC system initially increases and then decreases due to the trade-off between bandwidth and receiving area of the PD at the Rx, which limits the receiving power. At much higher bandwidths, the limitations on receiving power become more significant for CF than the contributions of bandwidth to the data rate. In contrast, for TeraCom systems, although an increase in bandwidth leads to higher noise levels, the contribution of bandwidth to the data rate is dominant in determining CF. However, enhancing the energy concentration of VCSEL beams will significantly help mitigate PD bandwidth limitations. In evaluations of link length impact on CF, we only considered the LoS components in THz channels for simplicity. Performance degradation occurs in both systems as link lengths increase, with the VCSEL-based OWC exhibiting greater sensitivity. While beam focusing improves performance, the VCSEL-based OWC is more suitable for short-range scenarios.

By integrating the performance evaluation models proposed for numerical analysis under misalignment conditions, we further illustrate the CF comparison when the Tx is tilted, displaced, or when both tilting and displacement are present, as in Fig.~\ref{f:cfcompmisaligned}. The results are obtained under the configuration that bandwidth $B=500$ MHz and link length $L=2$ m for both systems, $P_{\rm t}^{\rm thz}=40\mu$W for TeraCom system,  and $P_{\rm t}^{\rm v}=1$ mW for VCSEL-based OWC system. As in Fig.~\ref{f:cfcompmisaligned} (a) and (b), VCSEL-based OWC outperforms TeraCom in terms of CF under the same beam coverage conditions, and this advantage becomes more pronounced with increasingly focused beams. In Fig.~\ref{f:cfcompmisaligned} (c), we show CF as a function of Tx displacement distance with a tilt angle at $2^{\circ}$ at the same time. In conclusion, the impact of misalignment, specifically tilting and displacement, on VCSEL-based OWC performance arises from diffraction loss due to the shift between the beam area and the receiving area at the Rx side. In contrast, for TeraCom systems, the radiation patterns from both the Tx and Rx sides contributing to the overall antenna gain in a specific LoS direction, determine system performance under misalignment conditions.

\subsubsection{Coverage probability}
\begin{figure*}  
\centering
\subfigure[SNR threshold 5 dB]{
\begin{minipage}[b]{0.3\textwidth}
\includegraphics[width=1\textwidth]{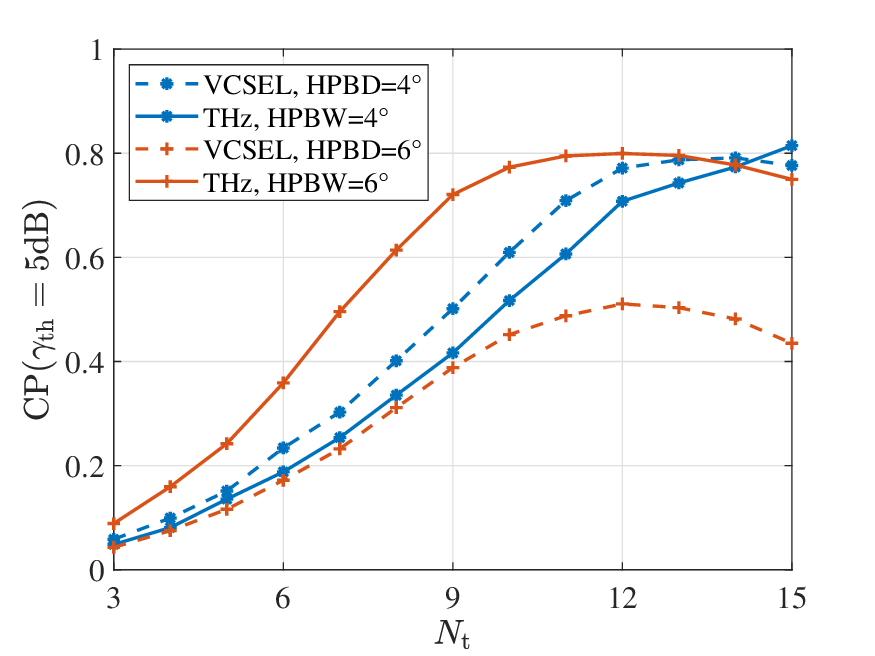}
\end{minipage}
}
\subfigure[$N_{\rm t}=12$]{
\begin{minipage}[b]{0.3\textwidth}
\includegraphics[width=1\textwidth]{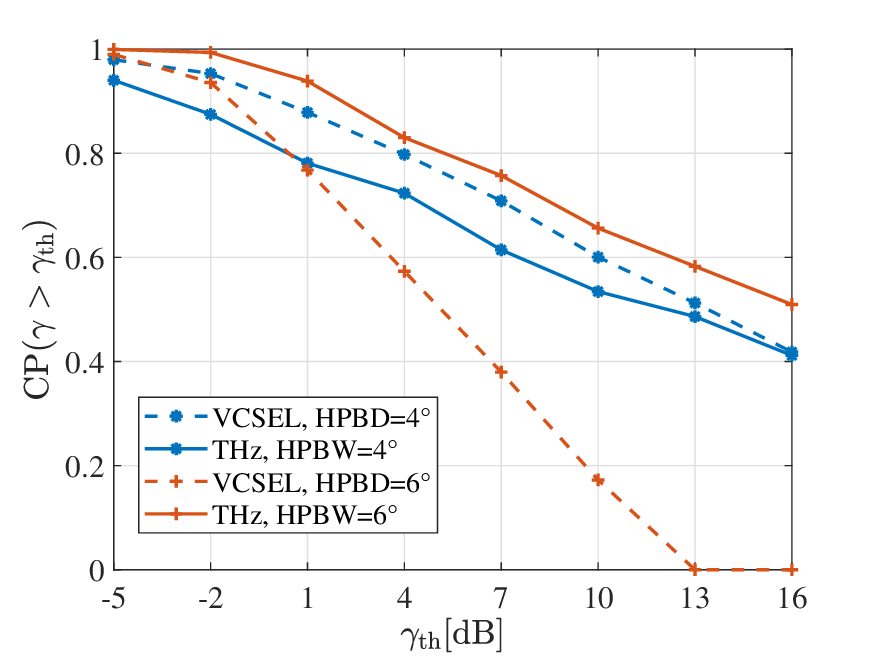}
\end{minipage}
}
\subfigure[$N_{\rm t}=12$]{
\begin{minipage}[b]{0.3\textwidth}
\includegraphics[width=1\textwidth]{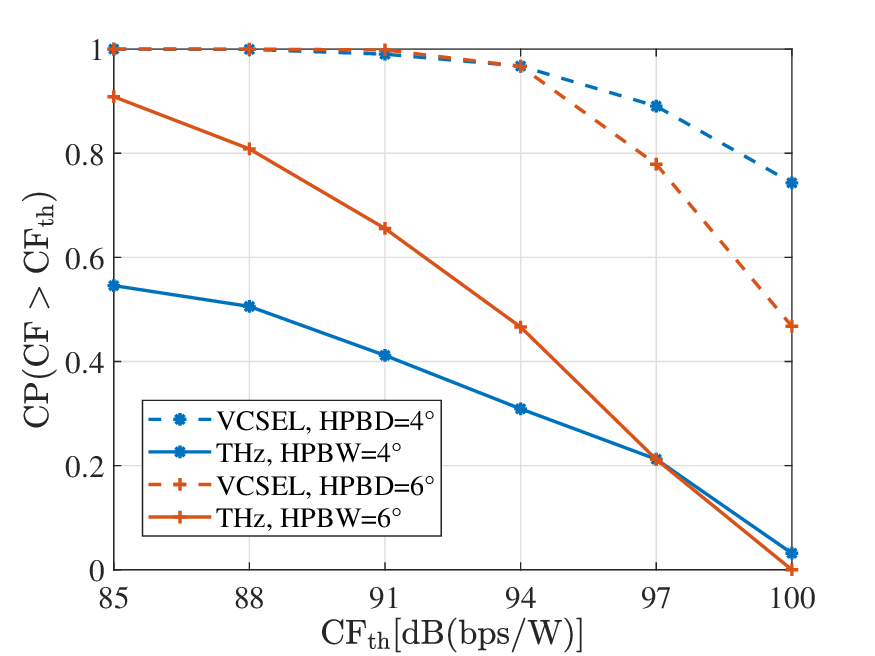}
\end{minipage}
}
\caption{Coverage probability in terms of SNR and CF.}
\label{f:cpcomp}
\end{figure*}
To verify the coverage probability of the two technologies with multiple transmitters mounted on the ceiling of the room, we analyze the following $3$ cases in Fig.~\ref{f:cpcomp}: i) probability of the SINR by a single user in arbitrary locations inside the room is larger than threshold $5$dB; ii) probability of SINR with varying threshold as $N_{\rm t}=15$; and iii) probability of CF with varying threshold as $N_{\rm t}=15$. The configurations are set as bandwidth $B=500$ MHz and link length $L=2$ m for both systems, $P_{\rm t}^{\rm thz}=40\mu$W for TeraCom system,  and $P_{\rm t}^{\rm v}=1$ mW for VCSEL-based OWC system. Fig.~\ref{f:cpcomp} (a) shows that as $N_{\rm t}$ increases, CP for both systems initially improves, reaching an optimal point before degrading due to interference noise when HPBW = HPBD = $6^{\circ}$. Overall, the CP for the TeraCom system is higher than that of the VCSEL-based OWC system. When HPBW = HPBD = $4^{\circ}$, CP increases as $N_{\rm t}$ improves from $3$ to $15$ for TeraCom system while in VCSEL-based OWC systems, there is still a saturation point. However, the VCSEL-based OWC system outperforms the TeraCom system, highlighting the importance of using a focused beam for optimal performance in VCSEL-based OWC systems.

From Fig.~\ref{f:cpcomp} (a), we choose $N_{\rm t}=12$ to validate the coverage probability with varying thresholds in terms of SINR and CF as in Figs.~\ref{f:cpcomp} (b) and (c), respectively. In Fig.~\ref{f:cpcomp} (b), both systems exhibit limited performance in a networked configuration, underscoring the critical need for beamforming or beam steering in this context. Still, beam focusing significantly impacts system performance regarding SINR, with the VCSEL-based OWC system being more reliant on it. At higher SINR thresholds, the VCSEL-based OWC cannot outperform TeraCom. However, as shown in Fig.~\ref{f:cpcomp} (c), the VCSEL-based OWC consistently outperforms in terms of CF, which is calculated solely based on the serving transmitter's power consumption. Besides, for TeraCom, a more focused beam does not necessarily lead to improved system performance.

\section{Outdoor applications: FSO vs. THz}
Free-space optical (FSO) communication and THz are envisioned to play key roles in outdoor applications, i.e., UAV-based, vehicular, and space communications. Channel models for the two technologies have been a research focus in analyzing system performance in outdoor application scenarios. In this section, we adopt widely used stochastic channel models for FSO and TeraCom to compare their performances in various application scenarios. After introducing the channel models, we specify the use cases of UAV, where we compare the two technologies quantitatively. 
\subsection{FSO stochastic channel model}
FSO channel gain consists of four impairments: i.e., atmospheric path loss $h_{\rm l}$, turbulence-induced fading $h_{\rm t}$, pointing errors $h_{\rm p}$, and angle of arrival (AoA) fluctuations induced link interruption $h_{\rm a}$; thus, the channel gain is given by
\begin{equation}
    h_{\rm fso} = h_{\rm l}h_{\rm t}h_{\rm p}h_{\rm a}.
\end{equation}
The atmospheric path loss for the FSO link is expressed below according to Beer’s-Lambert law
\begin{equation}
    h_{\rm l} = \exp(-\xi_{\rm l} L),
\end{equation}
where $\xi_{\rm l}$ is the attenuation factor closely related to the weather condition and $L$ is the link length. The effect of turbulence and misalignment are modeled by statistical processes, which are detailed in the following.
\subsubsection{Turbulence}
Málaga distribution is a general form as other commonly used distributions, e.g., Gamma-Gamma distribution is its special case. The probability density function (PDF) of the Málaga distribution which is used to model turbulence-induced fading is given by
\begin{equation}
    f_{h_{\rm t}}(h_{\rm t}) = A_M\sum_{s=1}^{\beta}a_sh_{\rm t}^{\frac{\epsilon+\beta}{2}-1}K_{\epsilon-\beta}\left(s\sqrt{\frac{\epsilon\beta h_{\rm t}}{g\beta+\Omega'}}\right),
\end{equation}
where the key factors are modeled as
\begin{equation}
    \left\{\begin{aligned}      
        &A_M  =\frac{2 \epsilon^{\epsilon / 2}}{g^{1+\epsilon / 2} \Gamma(\epsilon)}\left(\frac{g \beta}{g \beta+\Omega^{\prime}}\right)^{\beta+\epsilon / 2} \\
        &a_s  =\binom{\beta-1}{s-1} \frac{\left(g \beta+\Omega^{\prime}\right)^{1-s / 2}}{(s-1)!}\left(\frac{\Omega^{\prime}}{g}\right)^{s-1}\left(\frac{\epsilon}{\beta}\right)^{s / 2}\\
        &g = 2b_0(1-\rho),{\Omega}^{\prime}=\Omega+2 b_0 \rho+2 \sqrt{2 b_0 \rho \Omega} \cos \left(\varphi_a-\varphi_b\right)
    \end{aligned}\right.,
\end{equation}
where $K_{\epsilon-\beta}$ indicates the modified Bessel function of the second kind of $(\epsilon-\beta)$th-order, and $\Gamma(\cdot)$ indicates Gamma function. $\Omega, \varphi_a, \varphi_b$ represents the LoS component's average power, deterministic phase, and the deterministic phase of the coupled-to-LoS scatter component, respectively; $2b_0$ and $0<\rho<1$ denote average power of the total scatter components and amount of scattering power coupled to LoS component. $\epsilon$ and $\beta$ are large-scale and small-scale scattering parameters which rely on Rytov variance, link distance, and refractive index structure
parameter~\cite{scalepara}.

\subsubsection{Misalignment effect} The misalignment effect for the FSO link includes pointing errors and AoA fluctuation-induced link interruption. Considering the Gaussian beam profile at the receiving end, the PDF of pointing errors $h_{\rm p}$ is
\begin{equation}
    f_{h_{\rm p}}(h_{\rm p})=\frac{\xi_{\rm p}^2}{A_0^{\xi_{\rm p}^2}}h_{\rm p}^{\xi_{\rm p}^2-1},\quad 0\leq h_{\rm p}\leq A_0,
\end{equation}
where $\xi_{\rm p} = w_{\rm e}/(2\sigma_{\rm m})$ with $w_{\rm e}$ denoting the equivalent beam waist at the receiver side and $\sigma_m$ is the total displacement square variance. $A_0$ is the fraction of power collected at the detector without pointing errors. The above parameters are obtained as follows
\begin{equation}
    w_{\rm e}=\frac{w_{\rm z}^2 \sqrt{2} \operatorname{erf}(v)}{2 v \exp \left(-v^2\right)}, \quad A_0 =\operatorname{erf}(v)^2, \quad v = \sqrt{\frac{\pi}{2}} \frac{r_{\rm a}}{w_{\rm z}},
    \label{eq:A0}
\end{equation}
where $r_a$ is the aperture radius of the receiver, and $w_{\rm z}$ is the beam waist of the Gaussian beam at the receiver, which is given by~\cite{comprehensiveFSO}
\begin{equation}
    w_{\rm z} = w_0\sqrt{1+\left(1+\frac{2w_0^2}{(0.55C_n^2k_f^2L)^{-3/5}}\right) \left(\frac{L}{z_{\rm R}}\right)^2},
\end{equation}
where $C_n^2$ is the refractive index structure parameter and $k_f=2\pi/\lambda$ is the beam wave number.

Meanwhile, the PDF of AoA fluctuation is expressed as
\begin{equation}
    f_{h_{\rm a}}(h_{\rm a})=a_1\delta(h_{\rm a})+(1-a_1)\delta(h_{\rm a}-1), \quad a_1 = \exp \left(\frac{-\theta_{\rm FoV}^2}{a_2 \sigma_a^2}\right),
\end{equation}
where $\delta(\cdot)$ is the Dirac delta function,  $\theta_{\rm FoV}$ is the FoV of the receiver, $\sigma_a$ is the AoA fluctuation square variance, and $a_2=2$ or $a_2=4$ satisfy the situation where one or two sides have vibrations.
\subsubsection{Outage probability} Considering intensity modulation direct detection (IM/DD) for the FSO system and the above key parts in the FSO channel model, the outage probability is
\begin{equation}
    \begin{aligned}
 F_{\gamma_{\rm fso}}(\gamma_{\rm th}) \approx a_1+\left(1-a_1\right) D  \times \sum_{s=1}^\beta c_s G\begin{array}{l}
6,1 \\
3,7
\end{array}\left(A_1 \left\lvert\, \begin{array}{l}
1, K_1 \\
K_2, 0
\end{array}\right.\right), 
\end{aligned}
\end{equation}
where the details can be found as~\cite{comprehensiveFSO}
\begin{equation}\left\{
    \begin{aligned}
        &D=\xi_{\rm p}^2A_M/(8\pi), \quad c_s=2^{\epsilon+s-1}b_s,\quad b_s=a_s\left(\epsilon \beta /\left(g \beta+\Omega^{\prime}\right)\right)^{-(\epsilon+s) / 2},\\
        & A_1 = \frac{B^2\gamma_{\rm th}}{16\overline \gamma_{\rm fso}},~~B =\frac{\xi_{\rm p}^2 (g+\Omega')(1-a_1)\epsilon\beta}{(\xi_{\rm p}^2+1)(g\beta+\Omega')},\\
    &K_1=\left[{(\xi_{\rm p}^2+1)}/{2}, {(\xi_{\rm p}^2+2)}/{2}\right],\quad K_2 = \left[ \frac{\xi_{\rm p}^2}{2} , \frac{\xi_{\rm p}^2+1}{2}, \frac{\epsilon}{2} , \frac{\epsilon+1}{2}, \frac{s}{2}, \frac{s+1}{2}\right],
    \end{aligned}\right.
\end{equation}
where $\overline \gamma_{\rm fso}$ and $\gamma_{\rm th}$ denotes the average and threshold SNR of FSO link, and $G_{p,q}^{m,n}(\cdot)$ is the Meijer G-function.
\subsection{THz stochastic channel model}
THz channel gain includes the effect of path loss $g_{\rm l}$, multi-path fading $g_{\rm t}$, and misalignment $g_{\rm p}$, which is given by
\begin{equation}
    h_{\rm thz} = g_{\rm l}g_{\rm t}g_{\rm p}.
\end{equation}
As introduced in Eq.~\eqref{eq:losthz}, pass loss in THz link is described as $g_{\rm l}=\sqrt{G_0^{\rm Tx}G_0^{\rm Rx}}\frac{c}{4\pi fL}{\exp}[{\frac{1}{2}\kappa (f)L}]$, with $L$ indicating the link length. $\kappa (f)$ depends on relative humidity (RH). $G_0^{\rm Tx}$ and $G_0^{\rm Rx}$ denoting the maximum antenna gains at transmitter and receiver if there is no pointing error.
\subsubsection{Multi-path fading} $\alpha-\mu$ distribution is widely adopted to model the multi-path fading for THz links, where the PDF of $g_{\rm t}$ reads~\cite{alphamu}
\begin{equation}
    f_{g_{\rm t}}\left(g_{\rm t}\right)=\frac{\alpha \mu^\mu}{{\hat{g_{\rm t}}}^{\alpha \mu} \Gamma(\mu)} g_{\rm t}^{\alpha \mu-1} \exp \left(-\mu \frac{g_{\rm t}^\alpha}{{\hat{g_{\rm t}}}^\alpha}\right),
\end{equation}
where $\alpha, \mu$ denotes the fading parameter and normalized variance of the fading channel envelope, respectively, and $\hat{g_{\rm t}}$ is $\alpha$-root mean value of the fading channel envelope.
\subsubsection{Misalignment effect} For THz links, pointing errors are modeled with different methods. One widely used method is adopting the same pointing error model as in FSO link, where the misalignment effect is demonstrated by the power loss due to displacement of the THz Gaussian beam spot and THz receiver center. Then, the outage probability of the THz link is given by~\cite{thzmis}
\begin{equation}
F_{\gamma_{\rm thz}}(\gamma_{\rm th})=\frac{C_1 C_2}{\alpha}\left[\gamma_{\rm th}^{\frac{\xi_{\rm p}^2}{2}} / \bar{\gamma}_{\rm thz}^{\frac{\xi_{\rm p}^2}{2}}\right] \times G_{2,3}^{2,1}\left(C_{T_1} \gamma_{\rm th}^{\frac{\alpha}{2}} / \bar{\gamma}_{\rm thz}^{\frac{\alpha}{2}} \left\lvert\, \begin{array}{c}
1-\xi_{\rm p}^2 / \alpha, 1 \\
0, C_{T_2},-\xi_{\rm p}^2 / \alpha
\end{array}\right.\right)
\label{eq:thzbeam}
\end{equation}
Notations used in the expression are outlined by
\begin{equation}
    C_1=\frac{\xi_{\rm p}^2}{A_0^{\xi_{\rm p}^2}}, C_2=\frac{\mu^{\xi_{\rm p}^2 / \alpha}}{\hat{g_{\rm t}}^{\xi_{\rm p}^2} \Gamma(\mu)}, C_{T_1}=\frac{\mu}{A_{0}^\alpha \hat{g_{\rm t}}^\alpha},C_{T_2}=\frac{\alpha\mu-\xi_{\rm p}^2}{\alpha},
\end{equation}
where $\xi_{\rm p}$ and $A_0$ are modeled similarly to the FSO link, i.e.,  $\xi_{\rm p}$ is the ratio of equivalent beam waist to the displacement square variance and $A_0$ is represented in Eq. \eqref{eq:A0}.  

Recently, research showed that the misalignment effect model in THz links should be different from FSO links, where the antenna pattern is the key impact factor~\cite{generalTHzchannel}. The antenna gain is included in the pointing error model not the path loss model; thus with this derivation $g_{\rm l}'=\frac{c}{4\pi fL}{\exp}[{\frac{1}{2}\kappa (f)L}]$. The normalized pointing error is stated as
\begin{equation}
    g_{\rm p} = \sqrt{G^{\rm Tx}(\theta_{\rm ta}, \theta_{\rm te})}\sqrt{G^{\rm Rx}(\theta_{\rm ra}, \theta_{\rm re})},
\end{equation}
where $G^{\rm Tx/Rx}$ denotes the transceivers' normalized antenna radiation pattern with azimuth/elevation angle of antenna pointing direction at the transmitter side $\theta_{\rm ta}$,  $\theta_{\rm te}$, and antenna pointing angles at the receiver side $\theta_{\rm ra}$, $\theta_{\rm re}$. Characterizing the above variations as $\theta_{\rm ta}\sim \mathcal{N}(0, \sigma_{\rm ta})$, $\theta_{\rm te}\sim \mathcal{N}(0, \sigma_{\rm te})$, $\theta_{\rm ra}\sim \mathcal{N}(0, \sigma_{\rm ra})$, and $\theta_{\rm re}\sim \mathcal{N}(0, \sigma_{\rm re})$, we can define the pointing error factor vector as
$\overline{\xi_{\rm p}}=[\xi_1, \xi_2, \xi_3, \xi_4]=\left[\frac{\sigma_{\rm ta}^2}{(\theta_{\rm ta}^{\rm bw})^2}, \frac{\sigma_{\rm te}^2}{(\theta_{\rm te}^{\rm bw})^2}, \frac{\sigma_{\rm ra}^2}{(\theta_{\rm ra}^{\rm bw})^2}, \frac{\sigma_{\rm re}^2}{(\theta_{\rm re}^{\rm bw})^2}\right]$, where $\theta_{\rm ta,te,ra,re}^{\rm bw}$ indicate the HPBW of antennas in azimuth and elevation directions at transmitter and receiver sides, respectively. Then, the PDF of the pointing error is formulated as
\begin{equation}
f_{g_{\rm p}}\left(g_{\rm p}\right) \simeq C_g \frac{g_{\rm p}^{1 / \xi_{\rm p}'-1}}{G_0^{1 / \xi_{\rm p}'}} \sum_{k=0}^\infty \frac{\Delta_k\left(-\ln \left(\frac{g_{\rm p}}{G_0}\right)\right)^{k+1}}{\Gamma(k+2) \xi_{\rm p}'^{k+2}},
\end{equation}
with the assumption that the maximum antenna gain remains the same at the transmitter and receiver sides denoted by $G_0$ and $\xi_{\rm p}' = \min \{\overline{\xi_{\rm p}}\}$. With uniform linear array (ULA), antenna gain and HPBW are approximated by $\pi N_{\rm ant}^2$ and $1.061/N_{\rm ant}$, respectively, with $N_{\rm ant}\times N_{\rm ant}$ array antenna. Other notations are outlined as
\begin{equation}
    \left\{\begin{array}{l}
C_g=\prod_{i=1}^4 \sqrt{\xi_{\rm p}' / \xi_i},  \\
\Delta_k=\frac{1}{k} \sum_{i=1}^k i \xi_i \Delta_{k+1-i}~~\text { for } k=1,2,3, \ldots \\
\Delta_0=1, \quad \gamma_k=\sum_{i=1}^4 \frac{\left(1-\xi_{\rm p}' / \xi_i\right)^k}{2 k}
\end{array}\right.
\end{equation}
Then, considering both the multi-path fading and misalignment effect, the CDF of the THz link $h_{\rm thz}$ is modeled as
\begin{equation}
    \begin{aligned}
F_{h_{\rm thz}}(h_{\rm thz}) \simeq & 1-C_g \sum_{m=0}^{\mu-1} \sum_{k=0}^{\infty} \frac{A_2^m }{\Gamma(m+1)} \frac{\Delta_k \mathbb{A}_0(h)^m}{\Gamma(k+2) \xi_{\rm p}'^{k+2}}\times {\rm e}^{-A_2 \mathbb{A}_0(h)} \sum_{j=0}^{k+1} 2^{j-k-2}\binom{k+1}{j}  \\
& \left(A_1\right)^{k-j+1}  \times\left(\frac{A_3}{2}\right)^{\frac{j-3}{2}-k} \exp \left(\frac{A_1^2}{2A_3}\right)  \Gamma\left(\frac{j+1}{2}, \frac{A_1^2}{2A_3}\right)
\end{aligned}
\label{eq:thzant}
\end{equation}
with the key notations given by
\begin{equation}\left\{
\begin{aligned}
     &A_1=m \alpha-\frac{1}{\xi_{\rm p}'}-A_2 \mathbb{A}_0(h_{\rm thz}) \alpha,\quad A_2 = \frac{\mu}{\hat{g_{\rm t}}^{\alpha}}\\
    &A_3 = \alpha^2 A_2 \mathbb{A}_0(h_{\rm thz}),~\quad \mathbb{A}_0(h_{\rm thz}) = \left(\frac{h_{\rm thz}}{G_0g_{\rm l}'}\right)^{\alpha}
\end{aligned}\right.
\end{equation}
where $\Gamma(\cdot, \cdot)$ represents the upper incomplete gamma function. Denoting $h_{\rm thz} = g_{\rm l}'\sqrt{\frac{\gamma_{\rm th}}{\overline\gamma_{\rm thz}}}$, we can obtain the outage probability of THz link as $F_{h_{\rm thz}}\left(g_{\rm l}'\sqrt{\frac{\gamma_{\rm th}}{\overline\gamma_{\rm thz}}}\right)$.

\subsection{Comparison in UAV applications}
There are three types of links in UAV applications: UAV-to-UAV (U2U) where the transmitter and receiver apertures are installed on hovering UAVs, UAV-to-ground (U2G) where the information is transmitted from hovering UAV to ground base station, and ground-to-UAV (G2U), which operates in the reverse direction, from the ground station to the UAV~\cite{FSOUAV,FSOUAV2,comprehensiveFSO}. 
To analyze the FSO performance with the three above links considering the misalignment due to the movement of UAVs, the key is to model total displacement square variance $\sigma_{\rm m}$ and AoA fluctuation square variance $\sigma_{\rm a}$. With Cartesian coordinates, we denote $\sigma_{txp}, \sigma_{typ}$ as the standard deviation (SD) of transmitter position in $x-z$ and $y-z$ plane, $\sigma_{rxp}, \sigma_{ryp}$ as SD of the receiver position, $\sigma_{txp}, \sigma_{typ}$ as SD of the transmitter orientation, and $\sigma_{rxo}, \sigma_{ryo}$ as SD of the receiver orientation, respectively. Besides, non-zero foresight angle of the transmitter and receiver UAVs is considered, which is denoted by $\mu_{tx},\mu_{ty},\mu_{rx},\mu_{ry}$, respectively~\cite{FSOUAV,comprehensiveFSO}. Then, the total displacement variance for the three links is
\begin{equation}
    \begin{aligned}
\sigma_{\rm m}^2=  \begin{cases}\left(\frac{3 Z^2 \mu_{t x}^{ 2} \sigma_{d x}^4+3 Z^2 \mu_{t y}^{2} \sigma_{d y}^4+\sigma_{d x}^6+\sigma_{d y}^6}{2}\right)^{\frac{1}{3}}, & \text { U2U \& U2G } \\
\sigma_{t x p}^2+\sigma_{r x p}^2+\sigma_{t y p}^2+\sigma_{r y p}^2, & \text { G2U }\end{cases}
\end{aligned}
\end{equation}
where $\sigma_{d x}^2=L^2 \sigma_{t x o}^2+\sigma_{t x p}^2+\sigma_{r x p}^2$ and $\sigma_{d y}^2=L^2 \sigma_{t y o}^2+\sigma_{t y p}^2+\sigma_{r y p}^2$. AOA fluctuation variance is
\begin{equation}
    \sigma_a^2=\left\{\begin{array}{l}
\left(\frac{3 \mu_x^2 \sigma_x^4+3 \mu_y^2 \sigma_y^4+\sigma_x^6+\sigma_y^6}{2}\right)^{\frac{1}{3}}, \text { U2U }\\
\left(\frac{3 \mu_{t x}^{2} \sigma_{t x o}^4+3 \mu_{t y}^{2} \sigma_{t y o}^4+\sigma_{t x o}^6+\sigma_{t y o}^6}{2}\right)^{\frac{1}{3}}, \text { U2G }\\
\left(\frac{3 \mu_{r x}^{2} \sigma_{r x o}^4+3 \mu_{r y}^{2} \sigma_{r y o}^4+\sigma_{r x o}^6+\sigma_{r y o}^6}{2}\right)^{\frac{1}{3}}, \text { G2U } 
\end{array}\right.
\end{equation}
where $\mu_x=\mu_{tx}+\mu_{rx}$ and $\mu_y=\mu_{ty}+\mu_{ry}$. With model in Eq.~\eqref{eq:thzant} for UAV-based THz links, we use $\sigma_{\rm m}$ divided by link length to denote the total pointing angle error square variance.

\subsection{Comparison based on numerical analysis}
This subsection shows the FSO and THz link performance in outdoor environments and especially in UAV applications with exemplary system parameters. FSO link at $1550$ nm wavelength and THz link at $350$ GHz are chosen. For numerical analysis with above models, the commonly used parameters for FSO link are $\rho = 0.596$, $\Omega=1.3265$, $b_0=0.1079$, and for THz link are $\mu= 1$, 
$b_0= 0.1079$, $\hat{g_{\rm t}}=1$. The weather-dependent parameters for FSO link are from~\cite{comprehensiveFSO}.

\begin{figure*}  
\centering
\subfigure[FSO various weather]{
\begin{minipage}[b]{0.3\textwidth}
\includegraphics[width=1\textwidth]{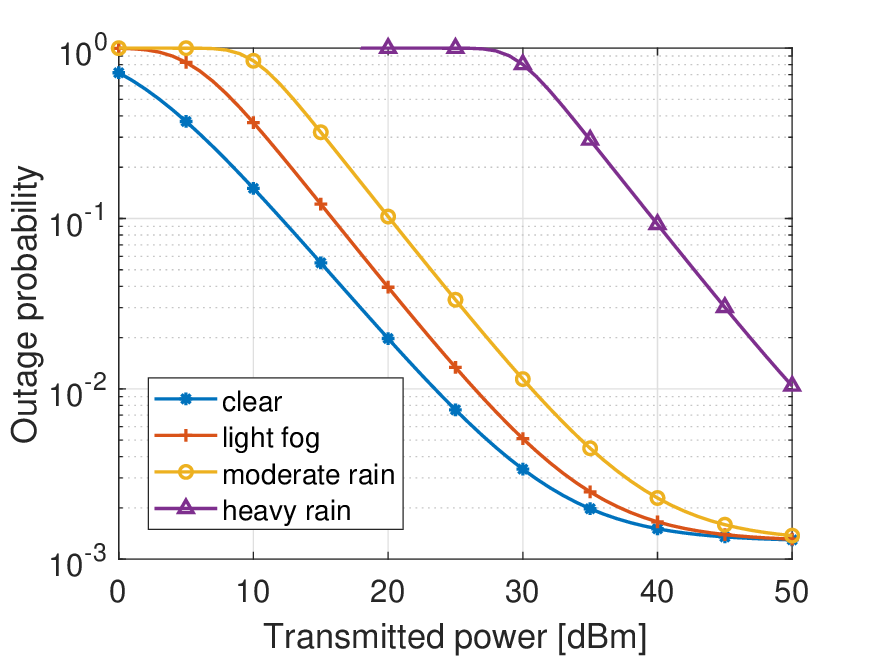}
\end{minipage}
}
\subfigure[THz humidity]{
\begin{minipage}[b]{0.3\textwidth}
\includegraphics[width=1\textwidth]{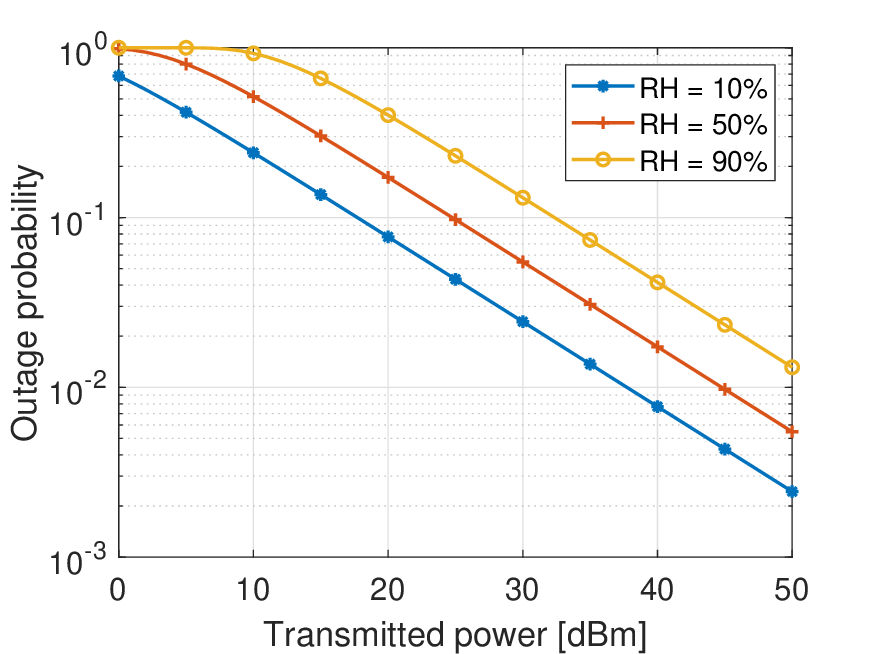}
\end{minipage}
}
\subfigure[THz spectral window]{
\begin{minipage}[b]{0.3\textwidth}
\includegraphics[width=1\textwidth]{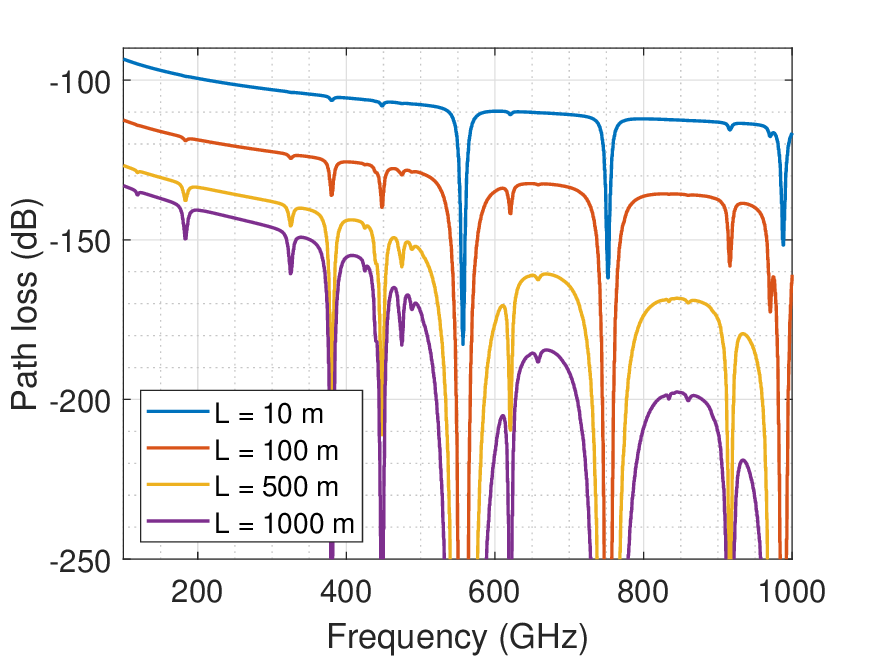}
\end{minipage}
}
\caption{FSO and THz links with various weather conditions, humidity, and frequency-dependent THz path loss.}
\label{f:stoweather}
\end{figure*}

\subsubsection{Impact of environmental conditions} Fig. \ref{f:stoweather} illustrates the performance of FSO and THz communication links under various environmental conditions and path loss characteristics. Here the outage probability indicates the probability of the instantaneous SNR falls below $5$dB. Model \eqref{eq:thzbeam} is used for THz link. For both links, $\omega_0 = 3$ m, $L = 1$ km, $\sigma_{\rm m} = 1.5$ m, radii of receiving plane $5$ cm, and noise power $-110$ dBm are set~\cite{pointingmodel}. For FSO link, $\theta_{\rm FoV} = 20$ mrad, responsivity equals $0.7$, and clear weather condition is assumed. For THz link, antenna gains at both transceivers are $55$ dBi and RH = $10\%$.

\begin{figure*}[t]  
\centering
\subfigure[THz model \eqref{eq:thzant} with antenna number]{
\begin{minipage}[b]{0.3\textwidth}
\includegraphics[width=1\textwidth]{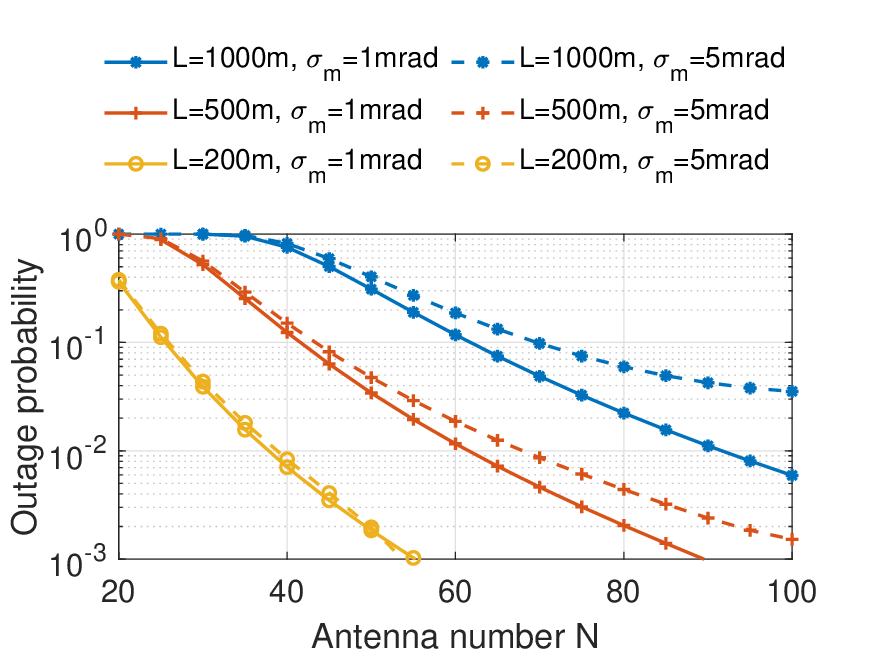}
\end{minipage}
}
\subfigure[THz model \eqref{eq:thzbeam} with beamwaist]{
\begin{minipage}[b]{0.3\textwidth}
\includegraphics[width=1\textwidth]{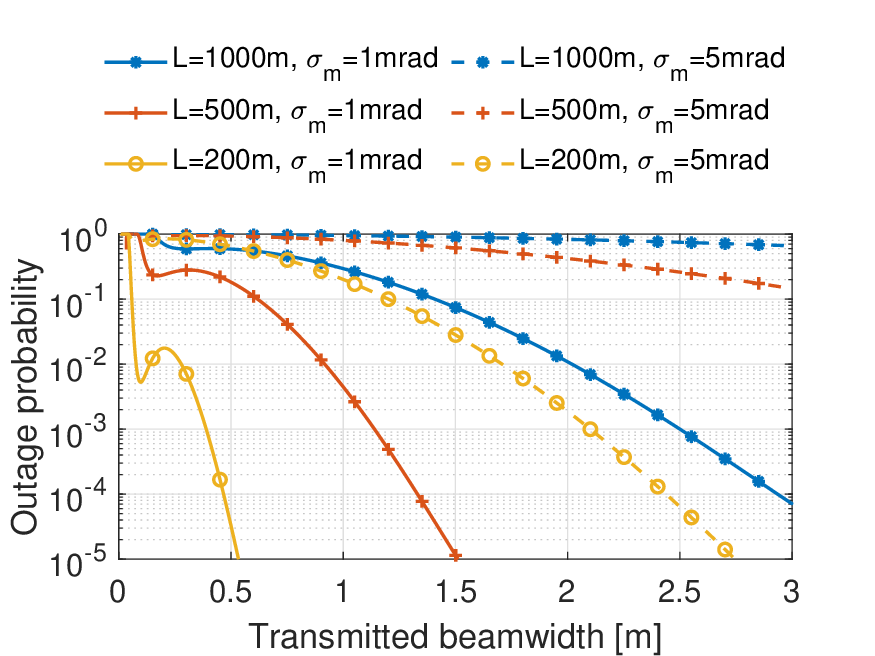}
\end{minipage}
}
\subfigure[THz model \eqref{eq:thzbeam} with pointing error]{
\begin{minipage}[b]{0.3\textwidth}
\includegraphics[width=1\textwidth]{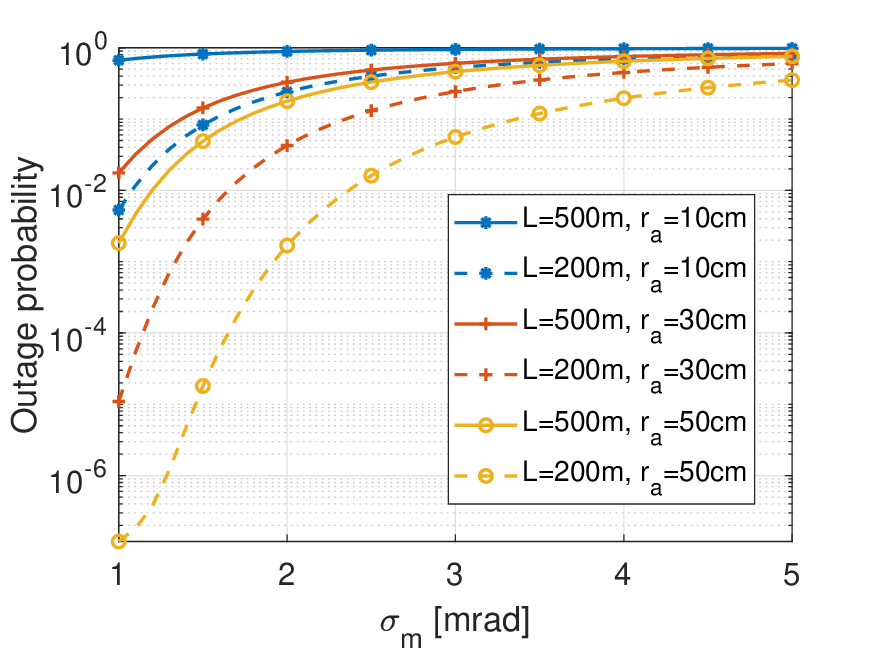}
\end{minipage}
}
\caption{Comparison of THz link with models \eqref{eq:thzbeam} and \eqref{eq:thzant}.}
\label{f:stoTHz}
\end{figure*}

\begin{figure*}[t]  
\centering
\subfigure[FSO]{
\begin{minipage}[b]{0.3\textwidth}
\includegraphics[width=1\textwidth]{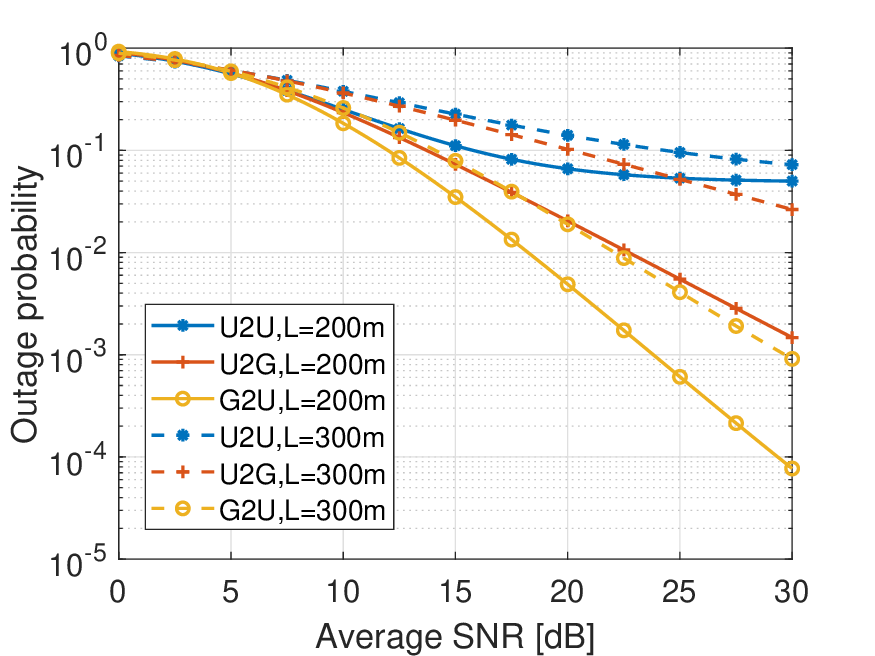}
\end{minipage}
}
\subfigure[THz]{
\begin{minipage}[b]{0.3\textwidth}
\includegraphics[width=1\textwidth]{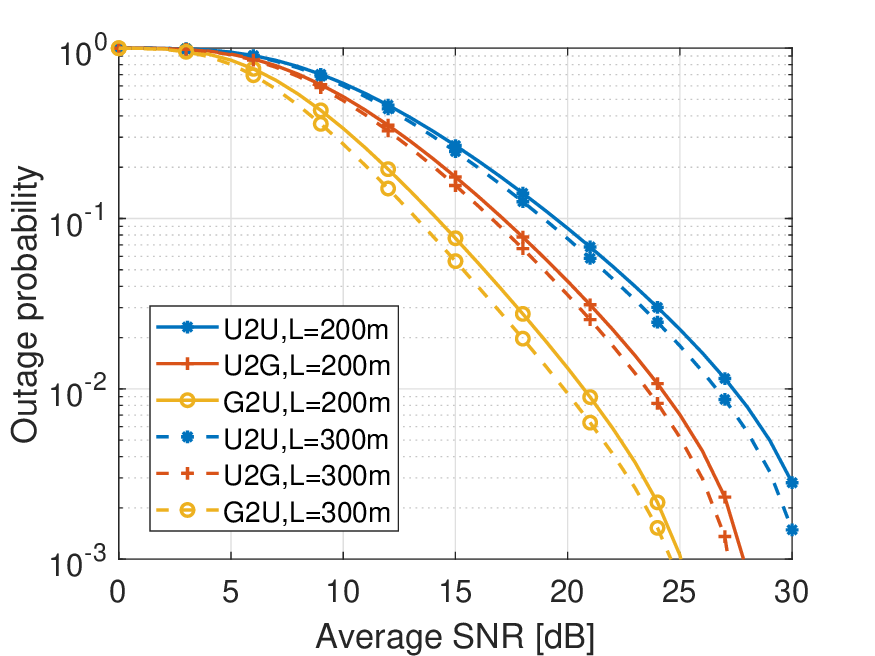}
\end{minipage}
}
\subfigure[Impact of THz antenna number]{
\begin{minipage}[b]{0.3\textwidth}
\includegraphics[width=1\textwidth]{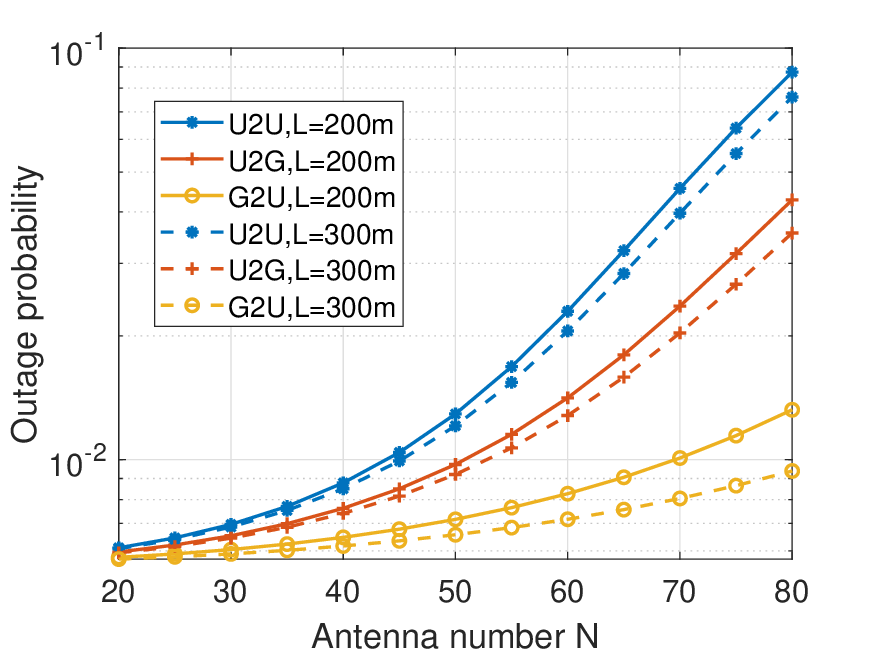}
\end{minipage}
}
\caption{Performance of FSO and THz links in UAV applications.}
\label{f:UAV}
\end{figure*}
As in Fig. \ref{f:stoweather} (a), the outage probability as a function of transmitted power for FSO links under different weather scenarios including clear weather, light fog, moderate rain, and heavy rain is depicted. As expected, adverse weather conditions (e.g., rain and fog) result in higher outage probabilities at lower transmission powers, with heavy rain having the most severe impact on signal reliability. Fig. \ref{f:stoweather} (b) presents the outage probability for THz communication links with varying relative humidity levels (RH = 10\%, 50\%, and 90\%). As the transmitted power increases, the outage probability decreases, but higher humidity levels (e.g., 90\% RH) cause significantly higher outage probabilities, indicating that THz link performance is sensitive to atmospheric humidity. Fig. \ref{f:stoweather} (c) further depicts path loss in dB across a range of frequencies ($0–1000$ GHz) for different transmission distances ($L$ = $10$ m, $100$ m, $500$ m, and $1000$ m). The graph reveals spectral windows where path loss is minimized, interspersed with frequencies experiencing high path loss due to MAL. Longer distances lead to greater path loss and narrow the effective spectral window, emphasizing the importance of frequency selection and distance in THz link design. Given the differing performance of FSO and THz links under various environmental conditions, a hybrid link design could effectively compensate for each of their limitations.

\subsubsection{Comparison of two THz link models} Fig.~\ref{f:stoTHz} presents a comparison of THz link performance using two models, Model \eqref{eq:thzant} and Model \eqref{eq:thzbeam}, considering the effects of antenna number, transmitted beamwidth, and pointing error. In this analysis, transmitted power is fixed at $5$ dBm with noise power $69$ dBm, and RH = $10\%$. Fig.~\ref{f:stoTHz} (a) shows the outage probability as a function of the number of antennas 
$N_{\rm ant}$, with different transmission distances $L$ ($1000$ m, $500$ m, and $200$ m) and pointing error standard deviations $\sigma_m$ ($1$ mrad and $5$ mrad). In model \eqref{eq:thzant}, the number of antennas determines the link gain and radiation HPBW, which influences the system's resilience to pointing errors. A higher number of antennas generally reduces outage probability, with shorter distances and smaller pointing errors further improving link reliability. On the contrary, in model \eqref{eq:thzbeam}, the impact of pointing error on THz system performance is determined by the receiving beam radius, similar to model for FSO link, as shown in Fig.~\ref{f:stoTHz} (b). The received beam radius depends on the transmitted beam radius according to the divergence principle of electromagnetic wave propagation. At the same time, a larger transmitted beam radius is aligned with the narrower HPBW, leading to higher antenna gain. Thus, larger beam radius with higher gain reduce the outage probability, especially at shorter distances. Fig.~\ref{f:stoTHz} (c) examines the impact of pointing error on outage probability for different receiver antenna radii $r_{\rm a}$ ($10$ cm, $30$ cm, and $50$ cm) and distances 
$L$, where $\omega_0 = 0.1$ m. As the pointing error standard deviation $\sigma_{m}$ increases, the outage probability rises, particularly for smaller receiver antenna radii and longer distances. Larger receiver antenna radii improve the system’s resilience to pointing errors, enhancing reliability over extended distances.

By comparing the two models, we found that model \eqref{eq:thzant} is less sensitive to link length. However, effective antenna area will also have an impact on the system performance considering the beam divergence, which is not demonstrated in model \eqref{eq:thzant}. Hence, the combination of two models to showcase full beam propagation characteristics is worthy of further investigation.
\subsubsection{Performance in UAV applications} Fig.~\ref{f:UAV} illustrates the performance of FSO and THz links in UAV applications, showing the outage probability for different link types (U2U, U2G, and G2U) over varying distances $L$ ($200$ m and $300$ m). We use model \eqref{eq:thzant} to present the performance of THz links, where for FSO links, $\omega_0 = 3$ m, $r_a = 5$ cm, $\theta_{\rm FOV} = 20$ mrad, and light fog weather condition is assumed. Correspondingly, RH=$40\%$ and $N_{\rm ant}=80$ are set for THz link. Primary parameters of misalignment for different links are from ~\cite{comprehensiveFSO}. Fig.~\ref{f:UAV} (a) shows the outage probability versus average SNR in dB for FSO links. Results indicate that G2U links have the lowest outage probability, followed by U2G links, and then U2U links. As the distance $L$ increases from $200$ m to $300$ m, outage probability increases across all link types due to higher path loss at longer distances. Similar to the FSO case, Fig.~\ref{f:UAV} (b) presents the outage probability as a function of average SNR for THz links. The same trend is observed where G2U links show superior performance, followed by U2G and U2U links. Higher average SNR levels lead to significantly reduced outage probability, especially for shorter distances (200 m) compared to longer ones (300 m). Through the comparison, we can find that FSO link is more sensitive to link length, and for U2U link with the largest pointing error probabilities, THz communications perform better. Fig~\ref{f:UAV} (c) examines the effect of the number of antennas $N_{\rm ant}$ on the outage probability for THz links. As the average SNR is fixed at $20$dB, increasing the antenna number primarily demonstrates the effect of narrower beamwidth, which increases the system’s sensitivity to pointing errors.

\section{Conclusion}
This study provides a quantitative assessment of THz and OWC technologies across indoor and outdoor environments, revealing key strengths and limitations of each. For indoor settings, our analysis shows that VCSEL-based OWC has higher energy efficiency than TeraCom but is more susceptible to distance and bandwidth limitations, requiring more focused beam transmission to address these challenges. In outdoor scenarios, stochastic channel models reveal that weather, absorption, and pointing errors significantly impact both FSO and THz links. FSO, with its heightened sensitivity to transmission distance and alignment, is less suitable for UAV-to-UAV applications compared to THz links. These findings underscore the importance of tailored technology choices depending on specific application needs and environmental conditions, providing guidance for selecting the optimal technology in both stable and dynamic environments.

\bibliographystyle{IEEEtran}
\bibliography{reference}
\end{document}